\documentclass{aa} 

\bibpunct{(}{)}{;}{a}{}{,} 

\usepackage[varg]{txfonts}
\usepackage{amssymb}
\usepackage{amsmath}
\usepackage{natbib}
\usepackage{subcaption}
\usepackage{graphicx}
\usepackage{epstopdf}
\usepackage{marvosym}
\usepackage{epsfig}
\usepackage{multirow}
\usepackage{tikz}
\usepackage{nicefrac}
\usepackage{arydshln}
\usepackage{url}
\usepackage{blkarray}
\usepackage{hyphenat} 
\usepackage{lipsum}
\usepackage{caption}
\usepackage{booktabs}
\usetikzlibrary{arrows,calc,intersections,positioning,arrows,decorations.pathmorphing,decorations.markings,shapes}

\newcommand{\RN}[1]{
	\textup{\uppercase\expandafter{\romannumeral#1}}
}

\begin{document}

\title{Combining astrometry and JUICE - Europa Clipper radio science to improve the ephemerides of the Galilean moons}

\author{M.~Fayolle \inst{\ref{TUDELFT}}\and A.~Magnanini\inst{\ref{BOLOGNA}}\and
V.~Lainey\inst{\ref{IMCCE}}\and D.~Dirkx\inst{\ref{TUDELFT}}\and M.~Zannoni\inst{\ref{BOLOGNA}}\and P.~Tortora\inst{\ref{BOLOGNA},\ref{BOLOGNA2}}}

\institute{Delft University of Technology, Kluyverweg 1, 2629HS Delft, The Netherlands \label{TUDELFT}\email{m.s.fayolle-chambe@tudelft.nl}
\and Dipartimento di Ingegneria Industriale, Alma Mater Studiorum - Universita di Bologna, 47121 Forli, Italy \label{BOLOGNA}
\and IMCCE, Observatoire de Paris, 77 Av. Denfert-Rochereau, 75014, Paris, France \label{IMCCE}
\and Centro Interdipartimentale di Ricerca Industriale Aerospaziale, Alma Mater Studiorum - Universita di Bologna, 47121 Forli, Italy \label{BOLOGNA2}
}

\date{Received 01 June 2023 / Accepted 17 July 2023}

\abstract 
{The upcoming JUICE and Europa Clipper missions targeting Jupiter's Galilean satellites will provide radio science tracking measurements of both spacecraft. Such data are expected to significantly help estimating the moons' ephemerides and related dynamical parameters (e.g. tidal dissipation parameters). However, the two missions will yield an imbalanced dataset, with no flybys planned at Io, condensed over less than six years. Current ephemerides' solutions for the Galilean moons, on the other hand, rely on ground-based astrometry collected over more than a century which, while being less accurate, bring very valuable constraints on the long-term dynamics of the system.} 
{An improved solution for the Galilean satellites' complex dynamics could however be achieved by exploiting the existing synergies between these different observation sets.}
{To quantify this, we merged simulated radio science data from both JUICE and Europa Clipper spacecraft with existing ground-based astrometric and radar observations, and performed the inversion in different configurations: either adding all available ground observations or individually assessing the contribution of different data subsets. Our discussion specifically focusses on the resulting formal uncertainties in the moons' states, as well as  Io's and Jupiter's tidal dissipation parameters.} 
{Adding astrometry stabilises the moons' state solution, especially beyond the missions' timelines. It furthermore reduces the uncertainties in $1/Q$ (inverse of the tidal quality factor) by a factor two to four for Jupiter, and about 30-35\% for Io. Among all data types, classical astrometry data prior to 1960 proved particularly beneficial. Overall, we also show that ground observations of Io add the most to the solution, confirming that ground observations can fill the lack of radio science data for this specific moon.} 
{We obtained a noticeable solution improvement when making use of the complementarity between all different observation sets. The promising results obtained with simulations thus motivate future efforts to achieve a global solution from actual JUICE and Clipper radio science measurements.}


\titlerunning{Global inversion for Galilean moons ephemerides}
\authorrunning{Fayolle et al}
\maketitle

\section{Introduction} \label{sec:introduction}

Due to the Laplace resonance between Io, Europa, and Ganymede, Jupiter's Galilean satellites form a complex dynamical system \citep[e.g.][]{lainey2006}. Reconstructing the long-term evolution of the Jovian system is thus extremely challenging, but will shed light on the formation and history of both the system itself \citep{peale1999,greenberg2010}, our own Solar System \citep{heller2015}, and exoplanetary systems in general \citep{horner2020}. In particular, an improved ephemerides' solution for the Galilean moons is expected to provide crucial insights into tidal dissipation mechanisms in the Jovian system, with direct implications for the moons' orbital evolution \citep{lainey2009,greenberg2010,fuller2016,hay2020}. This would also lead to a better characterisation of the moons' interior evolution, 
which would help to constrain the past and present properties of sub-surface oceans on Europa and Ganymede, as well as confirm the existence of such an ocean on Callisto \citep{spohn2003,schubert2004,greenberg2010,lunine2017}.

Current solutions for the Galilean moons' ephemerides \citep{lainey2004,lainey2009} mostly rely on ground-based astrometry, supplemented by space-based optical observations from Voyager and Galileo \citep{jacobson2000,haw2000,smith1979}. Radio science measurements acquired from the Galileo spacecraft during its moon flybys are included in a number of solutions, but the tracking accuracy was limited (S-band, low gain antenna) and not comparable to that of recent planetary missions \citep{jacobson2000,casajus2021}. Moreover, these data are not publicly available at present, limiting efforts to incorporate them into ephemeris solutions.

However, in the coming decade, the JUICE (JUpiter ICy moons Explorer) and Europa Clipper missions will both visit the Jovian system and specifically target the Galilean satellites. NASA's Europa Clipper mission will start its flyby tour in 2030 and perform more than 50 flybys of Europa. On the ESA side, the JUICE spacecraft will execute a series of flybys around the Galilean moons from 2032 to 2034 (two, seven, and nine at Europa, Ganymede, and Callisto, respectively). It will then enter an orbital phase of about eight months around Ganymede (first elliptical at an altitude of 5000 km, then circular at 500 km), before the planned mission end in 2035. The exceptional accuracy of the radiometric science data to be generated during the missions \citep[e.g.][]{cappuccio2022,mazarico2023}, enhanced by the complementarity of their tours and concurrent schedules, is expected to bring unique insights into the Galilean satellites' dynamics.

Several simulation studies have analysed the science return from both JUICE's and Clipper's radio science, and indicated that the post-missions formal uncertainties for the moons' state solutions could reach unprecedentedly low levels \citep{cappuccio2020, magnanini2021,fayolle2022}. However, exploiting the complementarity between the future radio science measurements and existing ground-based observations could yet further improve and stabilise the reconstruction of the Galilean moons' dynamics. For the Saturnian system, the independent determination of Titan's tidal dissipation parameters from ground astrometry and Cassini's radio science already led to very consistent solutions, indicating that Titan's migration rate might have been much faster than expected \citep{lainey2020}. This suggests the potential of global solutions capitalising on the synergies between diverse data types.

In practice, and for the Galilean moons in particular, astrometry and radio science are indeed very complementary datasets. While ground observations have been collected over more than a century, radio science measurements are by definition concentrated during planetary missions' timelines. Radio science thus provides shorter, but highly accurate data points. Existing ground-based observations are also relatively evenly distributed among the four Galilean satellites. JUICE and Europa Clipper, on the contrary, will provide an imbalanced dataset with a strong focus on Europa, Ganymede, and to a lesser extent Callisto, and no direct flyby performed at Io. The lack of data for Io has especially been identified as an important caveat in previous preparation studies for the two missions \citep[e.g.][]{dirkx2017}. The dynamics of Io, Europa, and Ganymede are indeed strongly coupled due to the Laplace resonance, such that  missing observational constraints for one of these three moons degrades the stability of the inversion and affects the estimation solution.

In this paper, we combine the existing astrometry and radar observations, used to generate the latest Galilean moons' ephemerides (NOE-5-2021\footnote{https://ftp.imcce.fr/pub/ephem/satel/NOE/JUPITER/2021/}), 
with radio science products from the JUICE and Europa Clipper spacecraft, as simulated in former analyses \citep{cappuccio2022,deMarchi2022,diBenedetto2021,magnanini2023,mazarico2023}.
We aim to quantify the synergy between the different datasets and its impact on the accuracy level of future ephemerides' solutions for the Galilean satellites. We considered different data subsets among all existing observations, and analyse their respective contribution to the joint solution. In particular, we investigated the improvement separately provided by classical astrometry, mutual phenomena, radar data, stellar occultations, and space-based astrometry. We finally quantified the added-value of potential future observation campaigns prior to JUICE's and Europa Clipper's Jovian tours.

On a more practical perspective, this study also demonstrates the ability to obtain a consistent solution while relying on several software currently tailored for different applications. The reconstruction of the moons' motion from astrometry is performed by NOE, a software developed in IMCCE (Institut de Mecanique Celeste et de Calcul des Ephemerides) used to generate state-of-the-art moons ephemerides for various systems \citep{lainey2009,lainey2016,lainey2019,lainey2020}. On the other hand, simulating JUICE and Clipper radio science observables and subsequently solving for both the spacecraft's and moons' dynamics was performed using two dedicated software packages: JPL's orbit determination software MONTE \citep[Mission Analysis, Operations, and Navigation Toolkit Environment,][]{evans2018} and Tudat (TU Delft Astrodynamics Toolbox). Various missions' radio science analyses already relied on MONTE \citep[e.g.][]{iess2018,durante2019,zannoni2020}, while Tudat, an open-source astrodynamics and estimation software developed at TU Delft, was used in a number of simulated estimation studies \citep[e.g.][]{dirkx2017,villamil2021,fayolle2021}. 

Both softwares were recently used to simulate the expected state solution for the Galilean moons from JUICE and/or Clipper radio science \citep{fayolle2022,magnanini2023} with slightly different estimation setups (Section \ref{sec:parameters}). In the present article, we retain the minor differences in estimation settings between the two tools, as we consider both to be equally representative of the estimation setup that will be used for the missions' data analysis. As will be shown in Section \ref{sec:results}, these minor differences in setup yield only minor differences in results. Keeping the small differences between MONTE and Tudat setups in our analysis not only provides important validation for the results provided by each tool, but it also provides more confidence in the robustness of the uncertainty results of one specific setup. 

All datasets used in our joint estimation are first presented in Section \ref{sec:dataSets}, starting with astrometry and radar observations before providing more details on the simulated JUICE and Europa Clipper radio science products. Section \ref{sec:inversionMethod} then describes the inversion strategy adopted in this work to combine not only different observation sets, but also different dynamical and propagation models for the spacecraft. The resulting global solution is discussed in Section \ref{sec:results}, along with detailed analyses of the contribution of different observation types, before conclusions can be drawn in Section \ref{sec:conclusion}.

\section{Datasets} \label{sec:dataSets}

This section describes the different observation sets to be included in the global inversion. The astrometry and radar data are first described in Sections \ref{sec:existingAstrometry} and \ref{sec:simulatedAstrometry}, followed by the simulated radio science measurements for both JUICE and Europa Clipper missions in Section \ref{sec:radioscience}. The synergies between the two datasets are finally further discussed in Section \ref{sec:synergies}.

\subsection{Existing astrometry} \label{sec:existingAstrometry}

\begin{figure*} [tbp!] 
	\centering
	\begin{minipage}[l]{1.0\columnwidth}
		\centering
		\subcaptionbox{\label{fig:observations1}}
		{\includegraphics[width=1.0\textwidth]{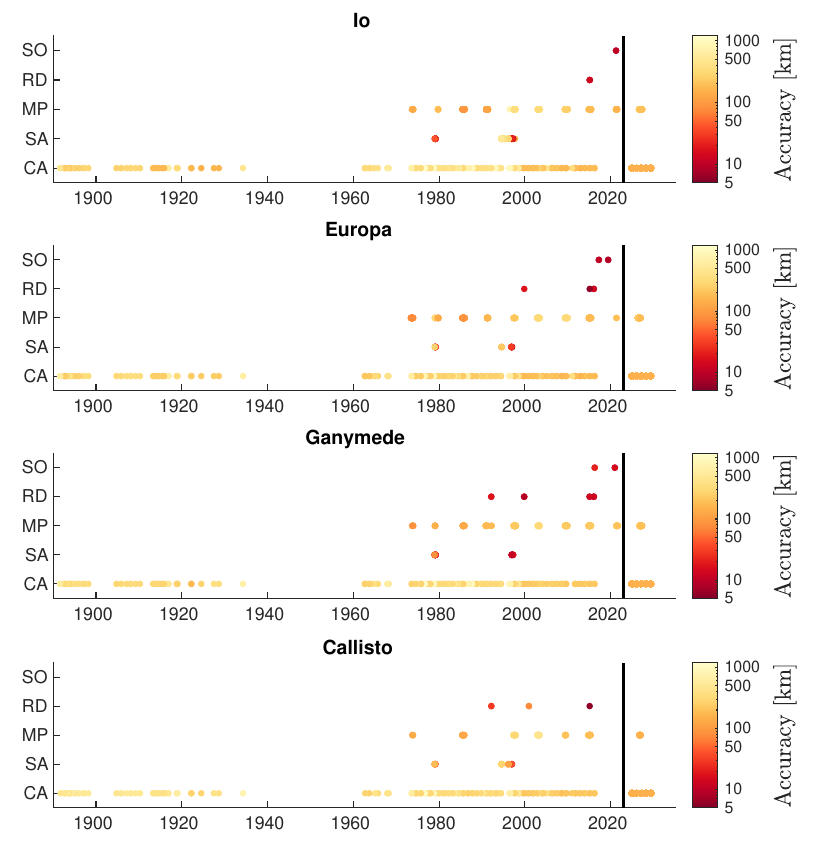}}
	\end{minipage}
	\hfill{} 
	\begin{minipage}[r]{1.0\columnwidth}
		\centering
		\subcaptionbox{\label{fig:observations2}}
		{\includegraphics[width=1.0\textwidth]{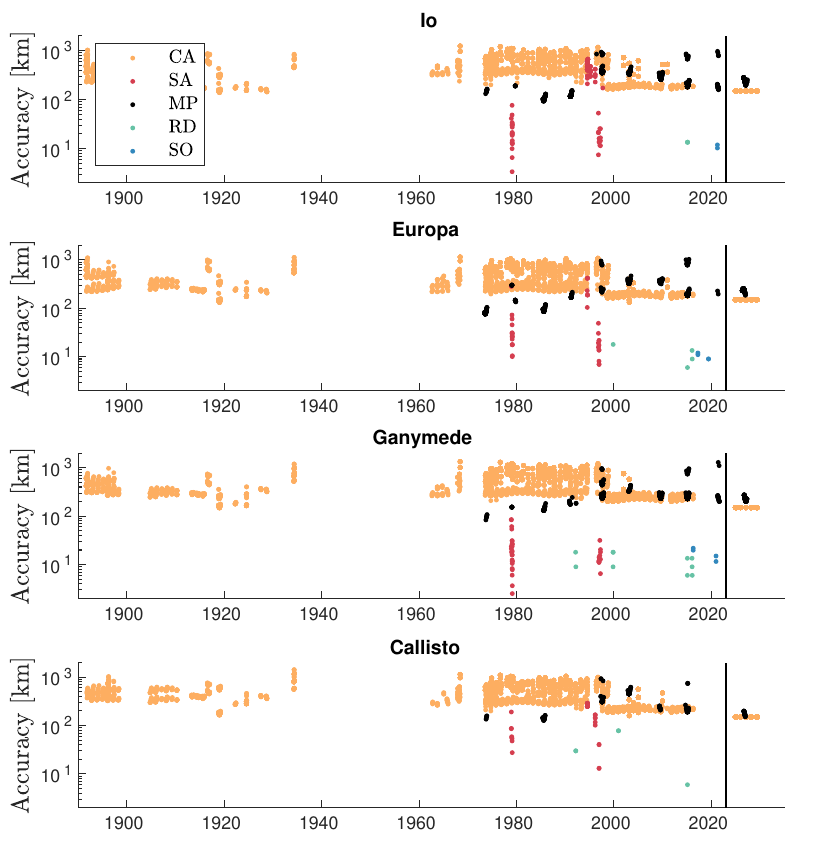}}
	\end{minipage}
	\caption{Properties of the ground- and space-based astrometry dataset, supplemented by a few radar observations. Panel a: ground- and space-based astrometric and radar data, divided into five main observation types: CA (classical astrometry), SA (space-based astrometry), MP (mutual phenomena), RD (radar), SO (stellar occultation). The colour indicates the accuracy of each measurement. Panel b: accuracy of the different ground- and space-based astrometric and radar data, for each of the main observation types. On both figures, the black vertical line delimits existing data and simulated future ones. It should be noted that the accuracy of the space-based astrometry is expressed with respect to the spacecraft, and not with respect to Earth as for the other observations.}
	\label{fig:observations}
\end{figure*}

All astrometric and radar observations used in IMCCE's latest ephemerides' solution for the Galilean satellites (NOE-5-2021) are shown in Fig. \ref{fig:observations}, the black vertical line separating past observations from predicted ones (after 2024). We distinguish between five main types of observations, displayed on separate rows. Astrometric data include both ground-based and space-based (SA) observations. The former encompasses classical astrometry (CA), mutual phenomena (MP), and stellar occultations (SO). Finally, existing radar observations of the Galilean satellites were also included in our dataset (RD). For the sake of brevity, all above observations are designated as the `astrometric dataset' in the rest of the paper even though they also include a few radar data.  

The weight assigned to each data point in the inversion (see Section \ref{sec:mergingPartials}), which can be interpreted as a measure of accuracy, is displayed on the vertical axis in Fig. \ref{fig:observations2} and colour-coded in Fig. \ref{fig:observations1}. These observation weights are nominally determined through an iterative process ensuring that they are consistent with the root-mean-square (RMS) of the residuals, and a $3\sigma$ ruling is applied to exclude outliers. Moreover, if many observations ($N$) are acquired in a short time span, such that they cannot be considered as independent measurements, they are de-weighted by a factor $\sqrt{N}$. More details on the weighting process for space-based astrometry in particular can be found in \cite{lainey2019}.

We include classical astrometry data from 1891 up to 2016. No data were available after 2016, as such observations are rarely performed for the Galilean satellites nowadays, due to the exceptional accuracy achieved with more novel observation techniques (e.g. stellar occultations (discussed below). Classical astrometry provides either absolute or inter-satellite position measurement in the plane of the sky. As shown in Figure \ref{fig:observations}, old astrometry typically shows low accuracy (several hundreds of kilometres). Nonetheless, some old photographic plates have been digitised and re-reduced using recent star catalogues \citep[e.g.][]{robert2011}, improving the accuracy of the observations. In addition to the data provided in \cite{robert2011}, recently reduced observations were also included in the estimation (not yet publicly available, from V. Robert, private communication).

Mutual phenomena designate occultations and eclipses of one moon by another. Such events require specific observation geometries, with the two moons aligned either with the Sun (eclipses) or with the Earth (occultations). For the Galilean satellites, they occur every six years when the Sun crosses their orbital plane. They have been observed since 1973, the latest mutual campaign to be recorded having taken place in 2021 \citep[e.g.][]{aksnes1976,arlot2006,arlot2014,emelyanov2022}.

Finally, stellar occultations currently represent the most accurate ground-based observation technique \citep{morgado2019,morgado2022} for the Galilean satellites. They rely on recording the drop in the photometric flux received by an observer as a moon passes in front of a star. With the help of recent Gaia star catalogues which provide a very accurate position for the occulted star \citep{gaia2018,brown2021}, the observation of the event allows to determine the moon's position in the ICRF (International Celestial Reference Frame) with an accuracy of a few milliarcseconds, equivalent to a few kilometres at Jupiter's distance. Since the availability of a highly accurate star catalogue is key to the quality of such observations, the first published stellar occultation by a Galilean moon (Europa) only occurred in 2017. Such events moreover require the Galilean satellites to pass in front of a bright enough star \citep[maximum magnitude of 11.5,][]{morgado2019}, and are therefore not very frequent. Only five stellar occultations are currently included in our dataset.

It is worth noting that the stellar occultations' uncertainty indicated in Fig. \ref{fig:observations} was actually artificially increased by 1.5 mas to account for the error in Jupiter's ephemeris \citep[e.g.][]{fienga2021}. This error source could eventually be mitigated using Gaia data, by extracting information about the Jovian system barycentre's position from Gaia's observations of Jupiter's outer satellites. This would be critical to achieve the expected few kilometres accuracy for stellar occultations. 

In addition to ground-based observations, space-based astrometry was also performed during planetary missions. In particular, both the Galileo and Voyager spacecraft were able to take images of the Galilean moons \citep[e.g.][]{haw2000,smith1979}. Those observations are interesting because of the different geometry under which they were taken, but they are affected by errors in the spacecraft orbit determination. For Galileo and Voyager, this error was very high compared to modern missions, significantly reducing the quality of the space astrometry data. In Fig. \ref{fig:observations}, their accuracy is expressed with respect to the spacecraft and not with respect to Earth, and thus cannot be directly compared with that of ground-based astrometric observations. 

Regarding ground-based radar observations, only 22 measurements from Arecibo are available \citep{brozovic2020}. Their number is limited, but they yield highly accurate measurements of the moons' line-of-sight ranges with respect to Earth (accuracy between 10 $\mu s$ and 250 $\mu s$ for the time delay measurement, equivalent to 6-80 km). The information provided by radar data actually distinguishes them from astrometric observations, which typically measure position(s) in the plane of the sky.

\subsection{Future astrometry for the period 2024-2029} \label{sec:simulatedAstrometry}

To complement the existing set of astrometric and radar data described in Section \ref{sec:existingAstrometry}, we also used future astrometric observations for the 2024-2029 period preceding the arrival of the JUICE and Europa Clipper spacecraft in the Jovian system, which were originally simulated for a past study investigating pre-mission ephemerides' solutions (V. Lainey, private communication). Including such synthetic data in our analyses allows us to quantify how much such Earth-based observations could contribute to a post-mission combined solution. Given the unprecedented accuracy level expected for the radio science products of both missions (see Section \ref{sec:radioscience} for more details), this is a key analysis to justify the need for future observation campaigns and identify which yet missing observations could efficiently complement JUICE's and Europa Clipper's data.

In addition to simulated classical astrometric observations, the upcoming mutual phenomena period, which will occur in 2027, is included. For classical astrometry, we generated around 1000 observations per moon and considered an accuracy of 150 km at Jupiter's distance, in agreement with the most accurate observations recently collected. This is representative of the expected accuracy for future observation campaigns, particularly given the availability of the very accurate Gaia GDR3 catalogue. Regarding mutual phenomena, 535 measurements were simulated (all moons combined), with accuracy levels comparable to recent observations (150-200 km). These synthetic observations are all reported in Fig. \ref{fig:observations}, on the right-hand side of the black line. As for radar observations, the set of existing data is limited to 22 measurements acquired since 1992. Given the loss of Arecibo, which dramatically reduces the ground-based radar observation capability, we chose not to include simulated radar data before the beginning of the JUICE and Europa Clipper missions. 

\subsection{Simulated radio science data} \label{sec:radioscience} 
The radio science dataset contains simulated range and Doppler measurements for the JUICE and Europa Clipper spacecraft. Both missions will generate such tracking data during flybys at Europa, Ganymede, and Callisto, as well as during Ganymede's orbital phase for JUICE. 

For JUICE's Jovian tour, we considered a X/Ka-band radio-link and assumed 48 hours of radiometric tracking centred at each flyby's closest approach, from the ESTRACK ground stations \citep{cappuccio2022}. We applied a noise of 20 cm for ranging measurements, which may be a rather conservative estimate given the recent performance of the BepiColombo radio science instrument \citep[sub-centimetre accuracy,][]{cappuccio2020,genova2021}, and 12 $\mathrm{\mu m/s}$ for Doppler data (at an integration time of 60 s). The orbital phase around Ganymede was divided into 24 hours-long arcs, with eight hours of tracking per day during which both range and Doppler measurements are generated. 

We simulated four hours of tracking at each closest approach for the Europa Clipper spacecraft, assuming a noise level of 0.1 mm/s as the X/Ka-band high-gain antenna (HGA) is then not available due conflict with other instruments \citep{mazarico2023}. The navigation tracking passes were however also included, in agreement with the current mission operation plan and recommended tracking setup for Clipper simulations \citep[e.g.][]{magnanini2023}. During these tracking arcs occurring further before or after the closest approach, the HGA can be used and the noise for Doppler data is thus divided by two. The average duration of the navigation tracking passes is five hours and they typically occur 20h before and after the closest approach. Range measurements could also be collected during such arcs, for which we assumed a noise of 1 m. 

\subsection{Synergistic combination} \label{sec:synergies}

Fig. \ref{fig:observations} highlights the main characteristics of the astrometry (and radar) dataset. Here, we described where synergies with radio science data will originate from. First, astrometry and radar observations are significantly less accurate than radio science measurements, with accuracies ranging from a few kilometres to several hundreds of kilometres depending on the observation type. However, they cover a much longer time span, starting in the 1890s until 2021, and even extending until the beginning of the JUICE and Europa Clipper missions if simulated data are included. Adding astrometry and radar observations is thus crucial to be sensitive to long-term signals in the moons' dynamics (e.g. dissipation effects). This is particularly important for the Galilean satellites as their dynamics show many long-period effects with different frequencies, which are difficult to distinguish from one another \citep{lainey2006}. Radio science measurements, on the other hand, are confined to the missions' timelines, for a total period of less than six years. The expected accuracy is however orders of magnitude better than what is achievable from ground-based observations \citep[e.g.][]{magnanini2023}. 

Furthermore, Fig. \ref{fig:observations} clearly illustrates that the astrometry dataset is more balanced than the radio science data, with observations more evenly distributed among the four Galilean satellites. JUICE and Europa Clipper, on the other hand, strongly focus on Europa and Ganymede, respectively. While Callisto is still targeted by a total of 30 flybys with both spacecraft, no flyby of Io is planned in the nominal mission scenarios. As already mentioned in Section \ref{sec:introduction}, adding existing astrometric and radar observations of this moon is thus particularly critical, since it is in mean-motion resonance with Europa and Ganymede. 

Finally, ground-based astrometry and radio science observables characterise the moons' dynamics under different observation geometries and are sensitive to the moons' motion projected in different directions. Astrometry typically measures the (absolute or relative) position of the satellites in the plane of the sky, while radio science's classical measurements, namely ranging and range-rate, probe the spacecraft's position and velocity in the Earth's line-of-sight direction. 

While comparing the astrometry and radio science datasets and their synergies, it is also worth noting a few of their major differences which might affect the estimation solution(s). First, astrometry and radio science measurements are affected by different noise sources. In particular, astrometry and radar data are sensitive to the offset between the centre of figure (COF) and centre of mass (COM), measuring the former while trying to solve for the latter. While this was not accounted for in our analysis, combining radio science with astrometry will be an effective way to estimate the COF-COM offset and thus mitigate this error source (see Section \ref{sec:conclusion}).

Radio science tracking, on the other hand, only indirectly probes the moons' dynamics around Jupiter, by reconstructing the spacecraft's trajectory as it passes in the close (gravitational) proximity of the Galilean satellites. This implies that additional parameters influencing the spacecraft's orbit determination solution need to be solved for concurrently with the moons' dynamics \citep{fayolle2022,magnanini2023}, as listed in Section \ref{sec:parameters}. In practice, the number of estimated parameters significantly increases when introducing radio science measurements, especially since all spacecraft-related parameters are typically solved for locally, in an arc-wise manner. This can affect the stability of the inversion and the estimation solution.

\section{Inversion methodology} \label{sec:inversionMethod}

\begin{figure*}[ht!]
	\centering
	\makebox[\textwidth][c]{\includegraphics[width=1.1\textwidth]{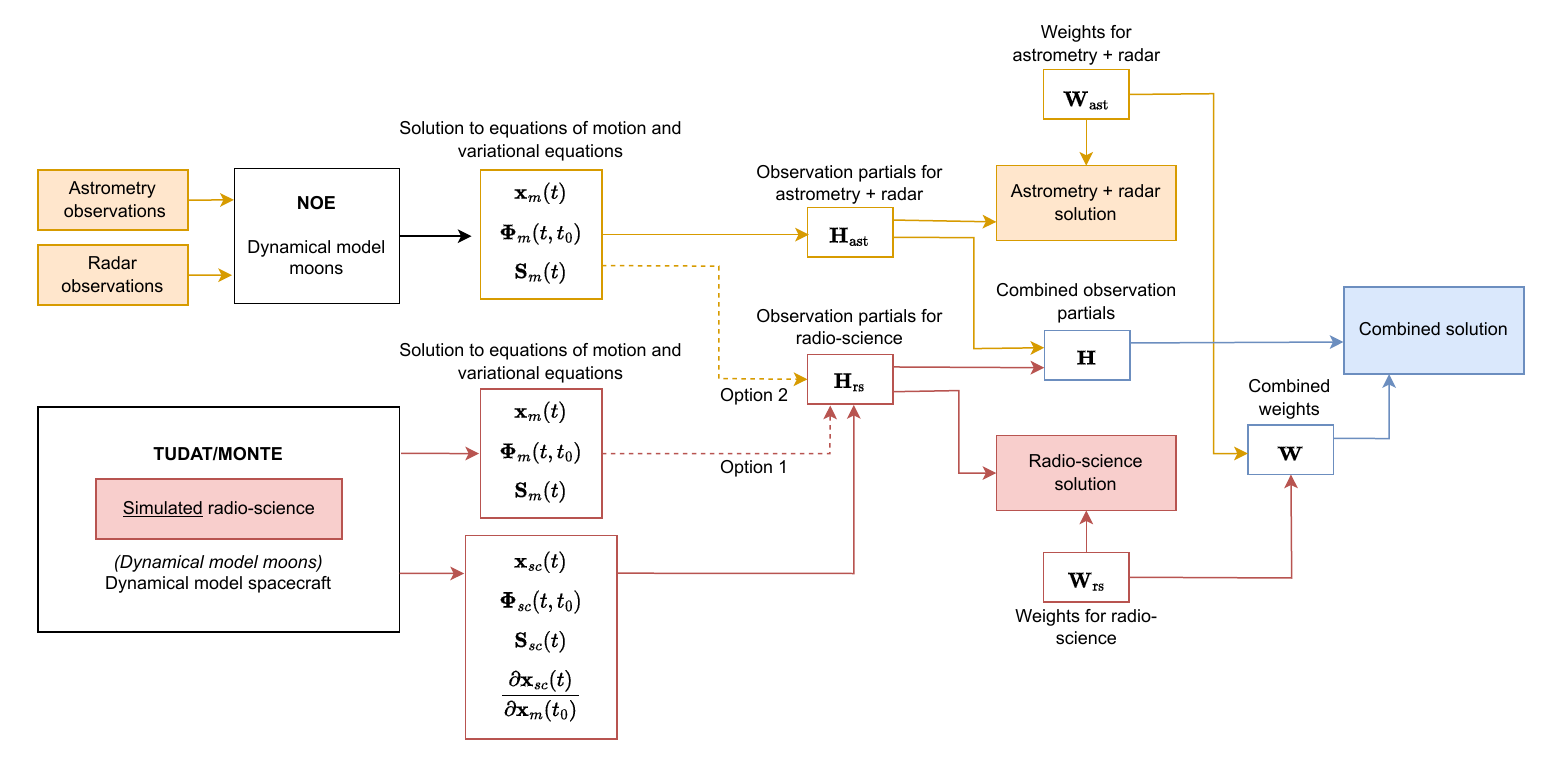}}
	\caption{Schematic summary of the global inversion methodology detailed in Sections \ref{sec:setups} and \ref{sec:mergingPartials}. The main interfaces between the three software (NOE, Tudat, and MONTE), as well as their inputs and outputs are represented. In particular, the two different strategies to include the solutions of the moons' equations of motion and variational equations in the radio science partials (Section \ref{sec:mergingPartials}) are illustrated by the dotted lines. The mathematical notations correspond to those used in Section \ref{sec:mergingPartials}.}
	\label{fig:inversion}
\end{figure*}

This section presents the adopted strategy to perform the global inversion of astrometric and radio science data. Section \ref{sec:setups} provides a top-level description of the propagation and estimation setups, before Section \ref{sec:mergingPartials} details the merging process to combine astrometry and radio science data in the estimation. Finally, the list of estimated parameters can be found in Section \ref{sec:parameters}.

\subsection{Propagation and estimation setups} \label{sec:setups}

As mentioned in Section \ref{sec:introduction}, we rely on three different softwares to obtain a combined solution with astrometry and radio science. The NOE software is used to propagate the moons' dynamics over the entire time span of the astrometric dataset and compute the associated partials (see Fig. \ref{fig:inversion}). To this end, the gravity fields of Jupiter and the Galilean satellites are modelled by spherical harmonics expansions, extended up to degree and order two for the moons and including zonal coefficients up to degree ten for Jupiter. The moons' rotation is assumed to be synchronous, with the tidal bulge pointing towards the empty focus of the orbit \citep{lainey2019,lari2018}. Jupiter's rotation include precession and nutations terms, following the IAU model \citep{archinal2018}. To propagate the dynamics of the Galilean moons, the following accelerations are considered \citep{lainey2004,dirkx2016}: mutual spherical harmonics acceleration between Jupiter and each moon $i$, mutual spherical harmonics acceleration in-between the Galilean moons, tidal dissipation (using the formulation presented in e.g. \citealt{lainey2017,lari2018}), third-body perturbation from Saturn and the Sun, and general relativity acceleration corrections.

On the radio science side, the equations of motion and variational equations for the spacecraft are solved with both Tudat and MONTE. Both tools propagate the dynamics of the spacecraft, using the latest JUICE\footnote{JUICE trajectory: juice\_mat\_crema\_5\_0\_20220826\_20351005\_v01: \url{https://www.cosmos.esa.int/web/spice/spice-for-juice}} and Europa Clipper \footnote{Clipper trajectory: 21F31\_MEGA\_L241010\_A300411\_LP01\_V4\_postLaunch\_scpse:  \url{https://naif.jpl.nasa.gov/pub/naif/EUROPACLIPPER/kernels/spk/}} trajectories as baselines. The gravitational accelerations exerted by the moons on the spacecraft are modelled using spherical harmonics gravity models up to degree and order two for Io, 13 for Europa, 50 for Ganymede, nine for Callisto. Additionally, the spherical harmonics gravitational acceleration from Jupiter (with zonal coefficients up to degree ten), point-mass gravitational accelerations from the Sun and Saturn, and the solar radiation pressure acceleration are considered. From the spacecraft's propagated trajectories, both softwares can simulate radio science measurements with expected noise levels (see Section \ref{sec:radioscience}) and provide the corresponding observation partials.

NOE and Tudat or MONTE thus each provide part of the required inputs, allowing us to perform a combined estimation (see Fig. \ref{fig:inversion}). Section \ref{sec:mergingPartials} describes in detail how the astrometric and radio science observations were merged in the inversion process. 

For the purpose of our analyses, we chose to primarily rely on covariance results (see details in Section \ref{sec:mergingPartials}). We indeed focus on quantifying the improvement attainable from a combined solution, which is well described by comparing formal uncertainties. It is however worth specifying that, as discussed in Section \ref{sec:existingAstrometry}, the weights assigned to the existing astrometric observations are based on the RMS of the residuals, and thus on real data analysis. 

\subsection{Merging astrometric and radio science partials} \label{sec:mergingPartials}

The inversion approach adopted in our analyses follows the coupled estimation strategy described in detail in \cite{fayolle2022}, with limited extensions to allow for the merging of different datasets, as summarised below. The covariance matrix $\boldsymbol{P}$ for the estimated parameters $\boldsymbol{p}$ is given by the following equation \citep[e.g.][]{montenbruck2002}:
\begin{align}
\boldsymbol{P} = \left(\boldsymbol{H}^{\mathrm{T}}\boldsymbol{W}\boldsymbol{H}+\boldsymbol{P}_0^{-1}\right)^{-1}, \label{eq:cov}
\end{align}
where $\boldsymbol{H}$ is the observations partial matrix, $\boldsymbol{W}$ is the matrix containing the weights to be applied to each observation and $\boldsymbol{P}_0$ is the a priori covariance matrix of the estimated parameters.

The full observation partial matrix $\boldsymbol{H}$ can be decomposed between the astrometric (denoted as ast) and radio science (rs) data subsets (see Fig. \ref{fig:inversion}), as follows:
\begin{align}
\boldsymbol{H} = \begin{pmatrix}
\boldsymbol{H}_{\mathrm{ast}}\\
\boldsymbol{H}_{rs}
\end{pmatrix} = \begin{pmatrix}
\frac{\partial \boldsymbol{h}_{\mathrm{ast}}}{\partial \boldsymbol{p}} \\[0.1cm]
\frac{\partial \boldsymbol{h}_{\mathrm{rs}}}{\partial \boldsymbol{p}}, 
\end{pmatrix}
\end{align}
where $\boldsymbol{h}_{\mathrm{ast}}$ and $\boldsymbol{h}_{\mathrm{rs}}$ represent the astrometric and radio science observations, respectively. The parameters vector $\boldsymbol{p}$ can be written as
\begin{align}
\boldsymbol{p} = \left[\begin{matrix}
\boldsymbol{x}_m(t_0) & \boldsymbol{x}_\mathrm{sc}(\boldsymbol{t}_i) & \boldsymbol{q}_\mathrm{dyn} &  \boldsymbol{q}_\mathrm{obs} 
\end{matrix}\right]^\mathrm{T}, \label{eq:parameters}
\end{align}
with $\boldsymbol{x}_m(t_0)$ the concatenated initial state vector for the four Galilean moons, and $\boldsymbol{x}_\mathrm{sc}(\boldsymbol{t}_i)$ the vector containing all the arc-wise initial states of both the JUICE and Europa Clipper spacecraft. $t_0$ refers to the global reference epoch, while $\boldsymbol{t}_i$ contains the initial times of each arc. $\boldsymbol{q}_\mathrm{dyn}$ and $\boldsymbol{q}_\mathrm{obs}$ correspond to the non-state parameters influencing the moons' and spacecraft's dynamics and the observations (e.g. biases), respectively. By definition, the astrometric observations are not sensitive to spacecraft-related parameters, such that only a subset of the parameters vector $\boldsymbol{p}$ is considered in the computation of their partials. In the merging process, zero-filled columns are thus added when relevant. 

Using Eq. \ref{eq:parameters} to expand the formulation for astrometric partials, we obtain for a single observation $h_\mathrm{ast}(t)$ in $\boldsymbol{H}_\mathrm{ast}$ ($t$ being the observation time):
\begin{align}
\frac{\partial h_\mathrm{ast}(t)}{\partial \boldsymbol{p}} &= \left[\begin{matrix}
\frac{\partial h_\mathrm{ast}(t)}{\partial \boldsymbol{x}_m(t_0)} 
& \frac{\partial h_\mathrm{ast}(t)}{\partial \boldsymbol{x}_\mathrm{sc}(\boldsymbol{t}_i)} 
& \frac{\partial h_\mathrm{ast}(t)}{\partial \boldsymbol{q}_\mathrm{dyn}} 
& \frac{\partial h_\mathrm{ast}(t)}{\partial \boldsymbol{q}_\mathrm{obs}} 
\end{matrix}\right]\\
&= \left[\begin{matrix} \frac{\partial h_\mathrm{ast}(t)}{\partial \boldsymbol{x}_m(t)}
\begin{pmatrix}
\boldsymbol{\Phi}_m(t,t_0)
& \boldsymbol{0}
& \boldsymbol{S}_m(t) 
\end{pmatrix}
& \frac{\partial h_\mathrm{ast}(t)}{\partial \boldsymbol{q}_\mathrm{obs}} 
\end{matrix}\right],
\end{align}
where $\boldsymbol{\Phi}_m(t,t_0)$ and $\boldsymbol{S}(t)$ are the state transition and sensitivity matrices for the moons dynamics, defined as:
\begin{align}
\boldsymbol{\Phi}_m(t,t_0) &= \frac{\partial\boldsymbol{x}_m(t)}{\partial\boldsymbol{x}_m(t_0)} \text{ };\text{ } 
\boldsymbol{S}_m(t)=\frac{\partial \boldsymbol{x}_m(t)}{\partial \boldsymbol{q}_\mathrm{dyn}}.
\end{align}

The astrometric subset of the design matrix $\boldsymbol{H}_\mathrm{ast}$ is thus fully derived from the variational equations solution for the moons' dynamics. On the other hand, the spacecraft's states are influenced by the moons' orbital motion, and the radio science partials $\boldsymbol{H}_\mathrm{rs}$ therefore depend on both the moons' and spacecraft's states:	
\begin{align}
\frac{\partial h_\mathrm{rs}(t)}{\partial\boldsymbol{p}} &=
\left[ \begin{matrix}
\frac{\partial h_\mathrm{rs}(t)}{\partial \boldsymbol{x}_m(t_0)} 
& \frac{\partial h_\mathrm{rs}(t)}{\partial \boldsymbol{x}_\mathrm{sc}(\boldsymbol{t}_i)} 
& \frac{\partial h_\mathrm{rs}(t)}{\partial \boldsymbol{q}_\mathrm{dyn}} 
& \frac{\partial h_\mathrm{rs}(t)}{\partial \boldsymbol{q}_\mathrm{obs}} 
\end{matrix}\right]\\
&= \left[\begin{matrix}
\frac{\partial h_\mathrm{rs}(t)}{\partial \boldsymbol{x}_\mathrm{sc}(t)}
\begin{pmatrix}
\frac{\partial\boldsymbol{x}_\mathrm{sc}(t)}{\partial\boldsymbol{x}_m(t_0)} 
& \boldsymbol{\Phi}_\mathrm{sc}(t,\boldsymbol{t}_i)
& \boldsymbol{S}_\mathrm{sc}(t) 
\end{pmatrix}
& \frac{\partial h_\mathrm{rs}(t)}{\partial \boldsymbol{q}_\mathrm{obs}}
\end{matrix}\right], \label{eq:partialsRs}
\end{align}
where $\boldsymbol{\Phi}_\mathrm{sc}(t,\boldsymbol{t}_i)$ and $\boldsymbol{S}_\mathrm{sc}(t)$ represent the state transition and sensitivity matrices for the spacecraft's dynamics. Because of the coupling between the moons' and spacecraft's dynamics, computing $\frac{\partial\boldsymbol{x}_\mathrm{sc}(t)}{\partial\boldsymbol{x}_m(t_0)}$ and $\boldsymbol{S}_\mathrm{sc}(t)$ in Eq. \ref{eq:partialsRs} also requires to solve the moons' variational equations \citep{fayolle2022}.

When computing the observation partials for (simulated) radio science data only with the MONTE or Tudat software, the equations of motion and variational equations for both the moons and the spacecraft are concurrently integrated, to account for the coupling in their dynamics. This implies that both NOE and Tudat or MONTE can provide their own solutions for $\boldsymbol{\Phi}_m(t,t_0)$ and $\boldsymbol{S}_m(t)$, as illustrated in Fig. \ref{fig:inversion}. Two different approaches can thus be considered:
\begin{enumerate}
	\item Independently integrating the moons' variational equations with different software;
	\item Importing the moons' solution provided by a single software into the other(s).
\end{enumerate}

The first option is straightforward to apply and allows for the direct stacking of the partials matrices $\boldsymbol{H}_\mathrm{ast}$ and $\boldsymbol{H}_\mathrm{rs}$, each generated with different software (NOE, and Tudat or MONTE, respectively) but using the same reference epoch $t_0$. However, the solutions to the equations of motion and variational equations for the moons must then be consistent between the two software, which was carefully verified in our case (between NOE and both Tudat or MONTE).

The second option is to directly import NOE's solution for $\boldsymbol{\Phi}_m(t,t_0)$ and $\boldsymbol{S}_m(t)$  into Tudat or MONTE to avoid propagating the moons' dynamics with different software. While this strategy is more demanding implementation-wise, it automatically ensures that the moons' dynamics are fully consistent between the astrometric $\boldsymbol{H}_\mathrm{ast}$ and radio science $\boldsymbol{H}_\mathrm{rs}$ partials. In practice, we implemented both options 
and showed that they indeed lead to equivalent results for our JUICE-Clipper case (see results in Section \ref{sec:results}). 

\subsection{Estimated parameters} \label{sec:parameters}
From the astrometric and/or radio science datasets, we estimate various parameters characterising the Jovian system and influencing the dynamics of the Galilean satellites. These include the initial states for Io, Europa, Ganymede, and Callisto, estimated at the reference epoch $t_0$ (set in the middle of the JUICE and Europa Clipper expected timelines), as well as the moons' gravitational parameters $\mu_i$, $i\in[1:4]$ and their gravity field coefficients up to degree and order 13, 50, and nine for Europa, Ganymede, and Callisto, respectively. Regarding Jupiter, we estimate its gravitational parameter $\mu_0$,  zonal coefficients ($J_2$ to $J_6$), and pole orientation (right ascension $\alpha$ and declination $\delta$ at the reference epoch $t_0$). Finally, tidal dissipation parameters include the $1/Q$ of Jupiter at a single frequency, and the $1/Q$ of Io at Jupiter's frequency.

Spacecraft-related parameters are also determined when including radio science in the solution. Different subsets of the following set of parameters are estimated in the MONTE and Tudat setups. In addition to the arc-wise initial states of the JUICE and Europa Clipper spacecraft (estimated in both setups), the spacecraft-related parameters include various observation-related parameters: range biases (both setups), antenna phase centre positions (MONTE only), accelerometer calibration factors (Tudat only), solar radiation pressure coefficients (MONTE only).

The above describes a simplified setup compared to detailed simulations studying the achievable radio science solution at the end of the JUICE and/or Clipper missions \citep{magnanini2021,magnanini2023,fayolle2022}. In particular, only Io's and Jupiter's tidal dissipation parameters are estimated, with no frequency-dependency introduced for $1/Q$. However, our study aims at quantifying the relative improvement achieved with global inversion with respect to a radio science or astrometry-only solution. Keeping the setup close to the one currently used for the astrometry-only inversion \citep[e.g.][]{lainey2009} facilitates this comparison and the analysis of the estimation results.

Furthermore, as shown by the list of estimated parameters, the nominal setups for the joint JUICE-Europa Clipper estimation in Tudat and MONTE show some small differences. These different setups were independently used in past radio science simulation studies \citep[][for MONTE and Tudat setups, respectively]{magnanini2023, fayolle2022}. The reason for keeping them as such is twofold. First, both are realistic and representative setups for simulation purposes, since the optimal estimation setup cannot be fixed before real data become available. Second, using perfectly identical setups is challenging due to the lack of certain software capabilities (e.g. antenna phase centre positions not readily available in Tudat) and/or to the significant modifications that it would have required (e.g. including accelerometer calibration factors in the MONTE setup). On the other hand, and especially considering the absence of a unique preconised setup for a joint radio science inversion of JUICE and Europa Clipper, keeping these discrepancies between the two softwares also allows us to verify that our general results are not affected by the details of the covariance analysis setup. 

It makes our comparative results more robust by ensuring that the specific settings chosen do not substantially impact the results, and are thus more representative in light of the potential deviations that are expected to arise between simulations and real data solutions. Section \ref{sec:results} will thus provide results obtained with both Tudat and MONTE. Simulated analyses of radio science experiments performed with different tools and slightly different setups should result in comparable results if both setups are representative. Experience from past missions shows that differences of  a factor two or three are not uncommon \citep[see for instance BepiColombo simulations in e.g.][]{schettino2015,imperi2018}. Moreover, a difference of a comparable order between absolute uncertainties in simulated analyses and real mission data analysis is to be expected.

\section{Results} \label{sec:results}

This section presents the results of the global inversion, in comparison with the astrometry only or radio science only solutions. We then quantify the individual contribution of various subsets of the astrometry dataset, distinguishing between different types of observations or different targets. A detailed comparison between JUICE-only and Clipper-only is provided in \cite{magnanini2023}.

\subsection{Combined solution from astrometry and radio science} \label{sec:combinedSolution}

\begin{figure*} [tbp!] 
	\centering
	\begin{minipage}[l]{1.0\columnwidth}
		\centering
		\subcaptionbox{\label{fig:decorrelations_rs}}
		{\includegraphics[width=1.0\textwidth]{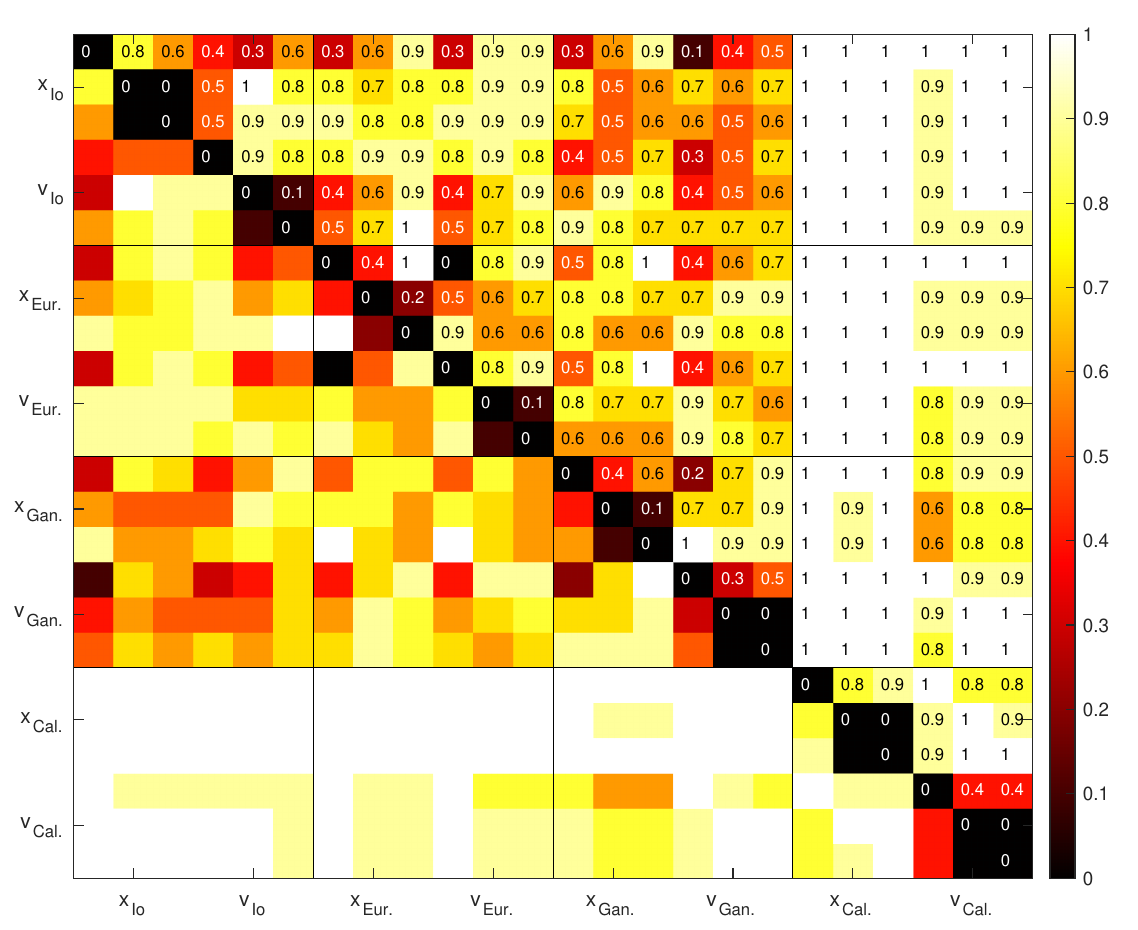}}
	\end{minipage}
	\hfill{} 
	\begin{minipage}[r]{1.0\columnwidth}
		\centering
		\subcaptionbox{\label{fig:decorrelations_change}}
		{\includegraphics[width=1.0\textwidth]{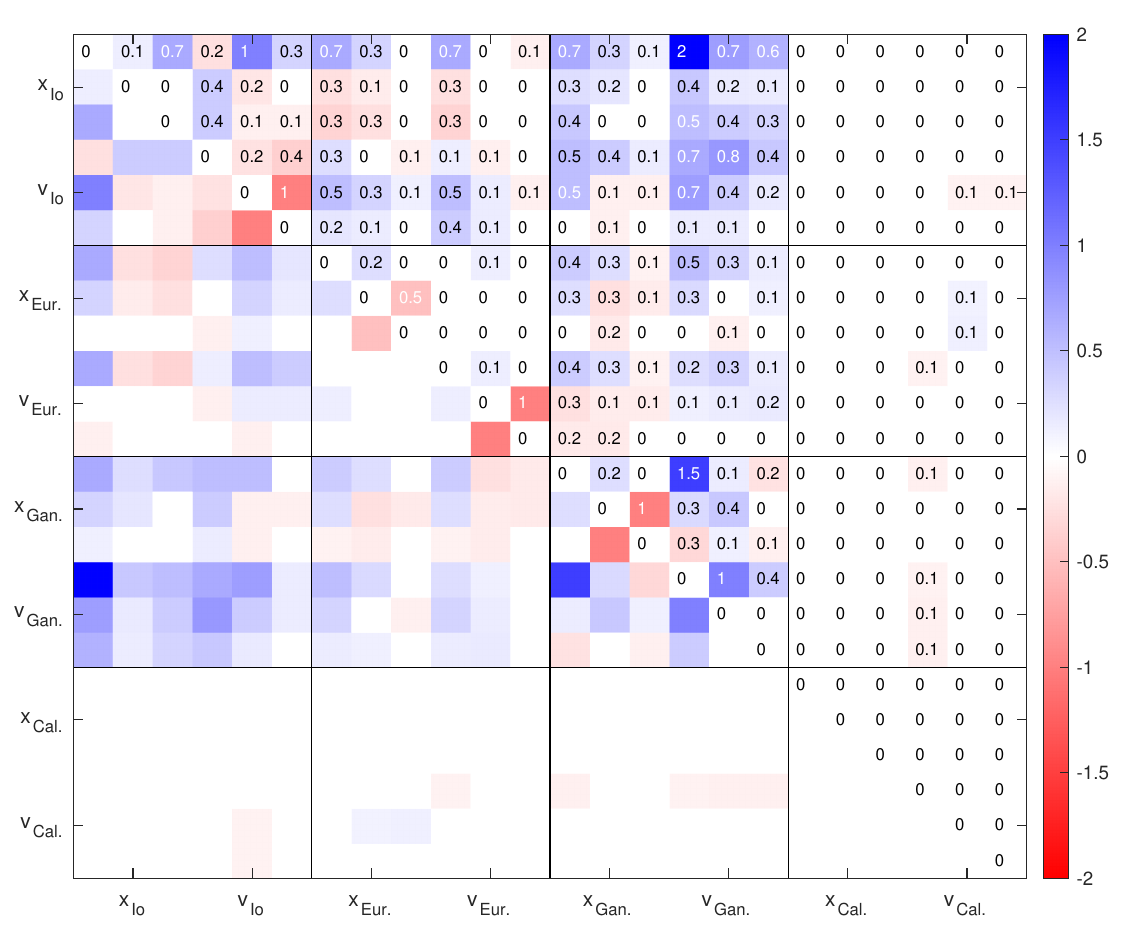}}
	\end{minipage}
	\caption{Effect of combining radio science and astrometry on the decorrelations between the Galilean moons' initial state components (in jovocentric cartesian coordinates). Panel a: decorrelations between the Galilean moons' states for the radio science-only solution. Darker colours indicate lower decorrelations between the parameters, thus stronger correlations. Panel b: Relative differences in decorrelations between the combined and radio science-only solutions (see Eq. \ref{eq:corr}). Blue and red indicate an increase and a decrease in decorrelation (i.e. a decrease and increase in correlation), respectively.}
	\label{fig:correlations}
\end{figure*}

\begin{figure*} [ht!]
	\centering
	\begin{minipage}[l]{1.0\columnwidth}
		\centering
		\subcaptionbox{\label{fig:errors_io_rs_missions_period}}
		{\includegraphics[width=1.0\textwidth]{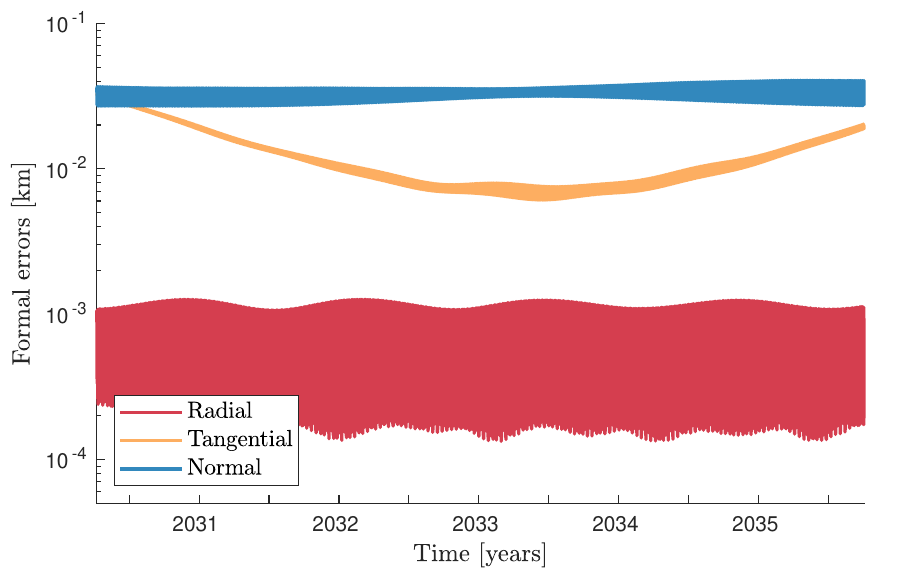}}
	\end{minipage}
	\hfill{} 
	\begin{minipage}[l]{1.0\columnwidth}
		\centering
		\subcaptionbox{\label{fig:errors_io_all_missions_period}}
		{\includegraphics[width=1.0\textwidth]{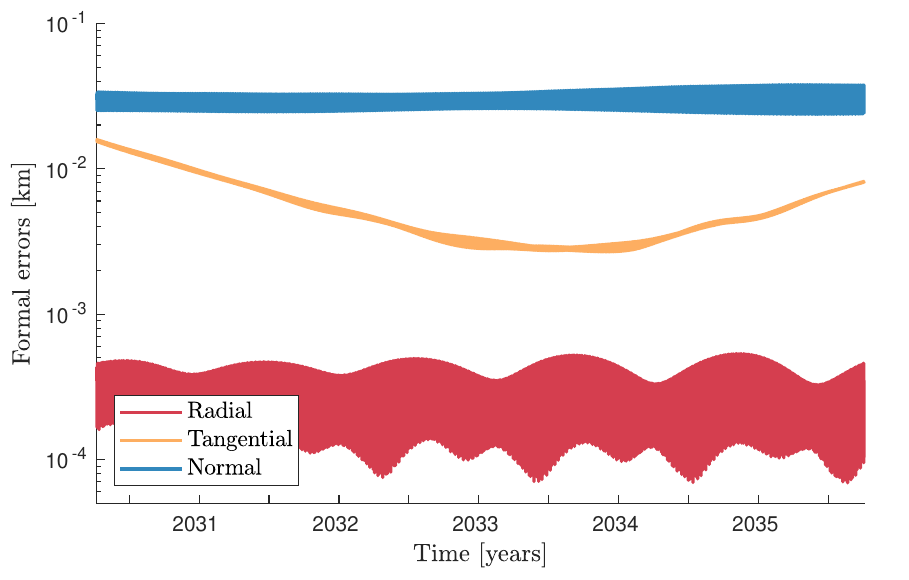}}
	\end{minipage}
	\caption{Evolution of the formal uncertainties in Io's state during the timelines of the JUICE and Europa Clipper missions. Panel a: radio science only solution, panel b: global inversion results (astrometry and radio science combined).}
	\label{fig:errors_io_missions_period}
\end{figure*}

As discussed in Section \ref{sec:introduction}, combining astrometric data with planetary missions' radio science measurements is mostly expected to help reconstructing the long-term orbital motion of the Galilean satellites. We thus focussed our analysis on the resulting uncertainties in the moons' states, as well as tidal dissipation parameters of Io and Jupiter. Three different simulated inversion solutions were generated: first with astrometry or radio science data only, and then using the complete observations set. All inversions were independently performed with both the Tudat and MONTE software, adding the astrometric observation partials and weights retrieved from NOE when relevant.

\subsubsection{Software consistency}

The solution based on astrometry only was used as a benchmark to compare the inversion results independently provided by our three software. This is to ensure that the computation of the covariance matrix according to Eq. \ref{eq:cov} is fully consistent between the tools. For astrometric observations, all inputs to Eq. \ref{eq:cov} are indeed identical, as they are directly provided by the NOE software (Fig. \ref{fig:inversion}), with no need to include the JUICE and Europa Clipper spacecraft in the estimation. The three solutions were in agreement and provided the same formal uncertainties for all estimated parameters. 

\begin{figure*} [tbp!] 
	\centering
	\begin{minipage}[l]{1.0\columnwidth}
		\centering
		\subcaptionbox{\label{fig:errors_europa_rs_full_period}}
		{\includegraphics[width=1.0\textwidth]{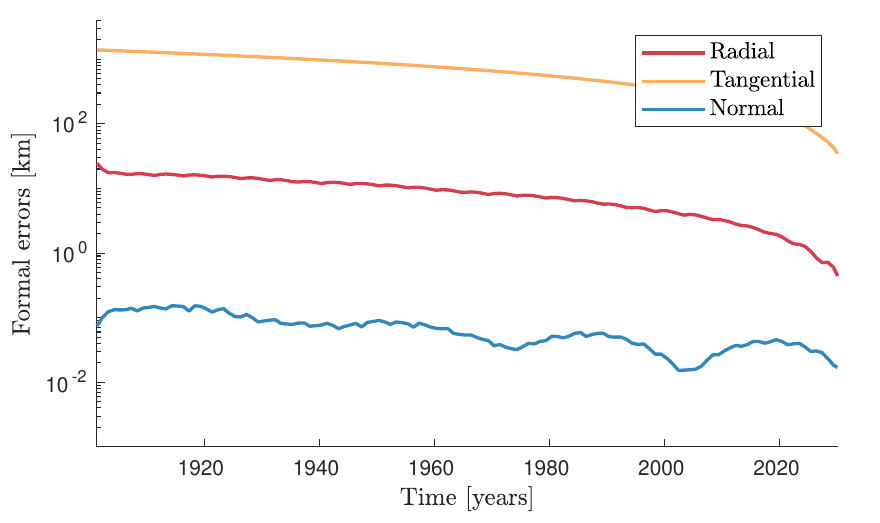}}
	\end{minipage}
	\hfill{} 
	\begin{minipage}[r]{1.0\columnwidth}
		\centering
		\subcaptionbox{\label{fig:errors_europa_all_full_period}}
		{\includegraphics[width=1.0\textwidth]{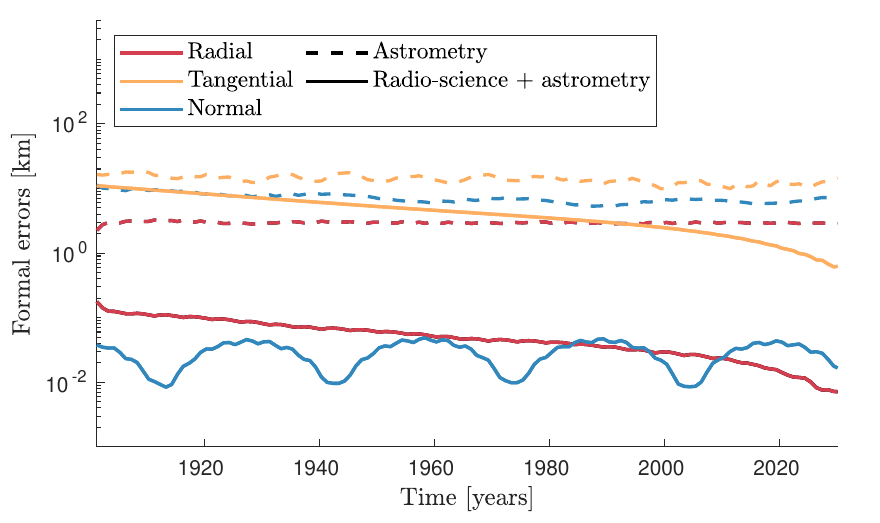}}
	\end{minipage}
	\caption{Propagated formal uncertainties in Europa's position (from 2030 to 1890) for the radio science, astrometry and combined solutions (obtained with Tudat and NOE). Panel a: radio science solution, panel b: astrometry-only and combined solutions. The errors are given in the RTN (radial, tangential, normal) directions and the scales are identical on both Fig. \ref{fig:errors_europa_rs_full_period} and \ref{fig:errors_europa_all_full_period}. To keep both the computational and memory loads manageable, we used a propagation output of one point per year and performed data smoothing over five-year windows to avoid aliasing effects. While this does not allow for local uncertainty analyses, it is nonetheless sufficient to investigate the long-term behaviour of the position errors far from the missions period.}
	\label{fig:errors_europa_full_period}
\end{figure*}

\begin{figure*}[ht!]
	\centering
	\includegraphics[width=1.0\textwidth]{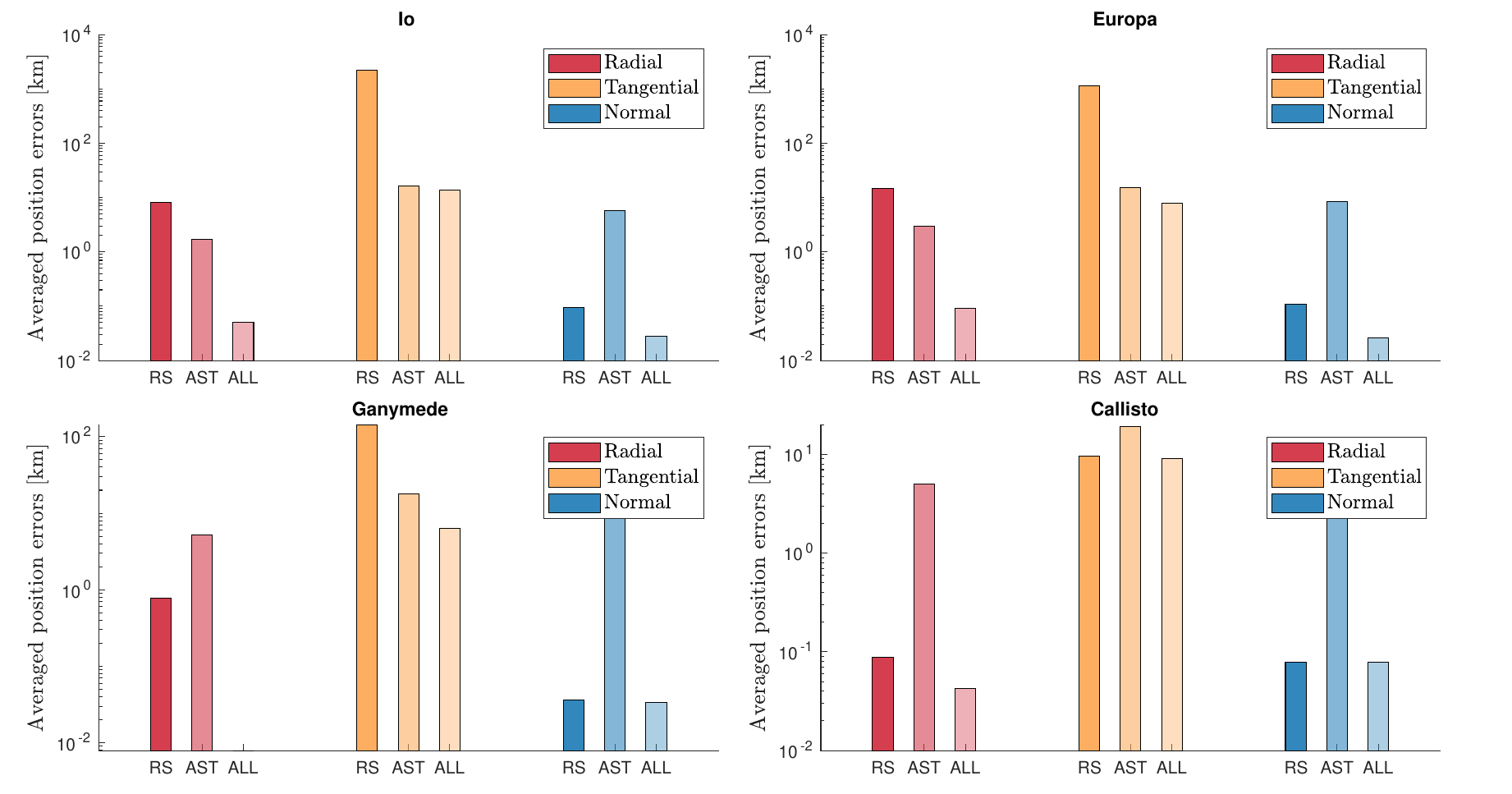}
	\caption{Formal uncertainties in the Galilean moons' satellites, averaged over the period 1890-2030. The results are provided for the radio science (RS), astrometry (AST), and combined (ALL) solutions.}
	\label{fig:averaged_errors}
\end{figure*}

For both the radio science and combined configurations, the two different approaches described in Section \ref{sec:mergingPartials} led to similar results (for a given software). For the sake of conciseness, Table \ref{tab:invQ} thus only provides one set of results for each software, which were equivalently obtained with both inversion strategies. Whether the solutions to the equations of motion and variational equations for the Galilean satellites were directly imported from the NOE software or separately recomputed when simulating the spacecraft dynamics did not affect the solution, demonstrating the consistency of the different software. In the rest of our analyses, the two approaches were thus considered equivalent. All Tudat results were obtained with both methods, while the second strategy was preferred in the MONTE setup (i.e. the moons' dynamical and variational equations were integrated with MONTE independently from NOE, for implementation reasons).

\subsubsection{Influence on the moons' state estimation}
\label{sec:resultsStateEstimation}
This section discusses the results of the global inversion for the Galilean satellites state estimation, analysing correlations (Fig. \ref{fig:correlations}) as well as formal uncertainties (Fig. \ref{fig:errors_io_missions_period}, \ref{fig:errors_europa_full_period}, and \ref{fig:averaged_errors}). First looking at the impact on the correlations between state parameters, Fig. \ref{fig:correlations} shows the relative change $\epsilon$ in absolute decorrelation when adding astrometry, with respect to the radio science only case, defined as follows:
\begin{align}
\epsilon = \frac{\left(1 - |c_{ij}^\mathrm{all}|\right) - \left(1 - |c_{ij}^\mathrm{rs}|\right)}{1 - |c_{ij}^\mathrm{rs}|}, \label{eq:corr}
\end{align} where $1-|c_{ij}^\mathrm{rs}|$ and $1-|c_{ij}^\mathrm{all}|$ are the decorrelation between parameters $i$ and $j$ in the radio science only and combined (radio science and astrometry) cases, respectively. Focusing on decorrelations (1 $-$ correlations) rather than correlations allows to scale changes as a function of the distance to full correlation: a decrease in correlation between two parameters indeed has a stronger influence on the inversion if the two parameters were originally fully correlated than if they were already rather decorrelated. 

Fig \ref{fig:correlations} shows that including astrometry in the solution decreases the correlations for most state parameters (shown as decorrelation increase in Fig. \ref{fig:correlations}). This is not only observed between state components of the same moon, but also between different moons. For fewer parameters, the correlations actually increase, which can be caused by the heterogeneous effect of adding more data (i.e. more information) on the uncertainties of different estimated parameters. If the additional observations help to reduce the uncertainty in parameter $i$, its correlation with parameter $j$ whose uncertainty remains unchanged might increase. Nonetheless, adding astrometry overall reduces the strong correlations between the states of Io, Europa, and Ganymede caused by the Laplace resonance. This improvement originates from adding observations over a longer time span, as well as direct measurements of Io's position which are critically missing in the JUICE and Europa Clipper radio science dataset.

Similar observations can be made from the state uncertainties obtained for the Galilean moons. During the period of the two missions, only Io's solution benefits from adding astrometry data to the estimation (Fig. \ref{fig:errors_io_missions_period}), while the other moons' states could not be improved in this time interval beyond the radio science solution. This is expected given the unprecedentedly low uncertainties predicted by simulations for JUICE and/or Europa Clipper \citep{fayolle2022,magnanini2023}. For a joint solution relying on both missions, the position uncertainties for Europa, Ganymede, and Callisto achieve sub-meter accuracy in the radial direction, while they do not exceed a few tens of meters in the tangential and normal directions. However, when using radio science data only, Io's state solution is solely based on indirect constraints originating from the well-determined dynamics of the other two moons in resonance. Including astrometry thus has a stronger effect for this moon: it reduces the averaged uncertainty in the radial and tangential positions by about a factor two (Fig. \ref{fig:errors_io_missions_period}). 

While radio science measurements alone already provide an extremely accurate solution for the Galilean moons' states during the JUICE and Europa Clipper missions, the formal uncertainties start to dramatically increase once propagated beyond the missions' timeline, especially in the radial and tangential directions. The solution is particularly unstable for the three moons in resonance whose dynamics are mutually affected by their state uncertainties. This is illustrated in Fig. \ref{fig:errors_europa_rs_full_period}, using Europa as an example (results for the other moons can be found in Appendix \ref{appendix}). The propagated uncertainties show a similar increase with time for both radial and tangential positions, although the errors in the radial direction remain about two orders of magnitude lower than the tangential ones. Uncertainties in the normal direction, on the other hand, do not strongly degrade upon propagation. These differences between in-plane and out-of-plane uncertainty propagation can be explained by the fact that most dynamical perturbations, as well as the Laplace resonance, act within the moons' orbital plane. This causes in-plane position errors to quickly propagate into larger uncertainties, which however do not strongly affect the out-of-plane motion. 

Merging the radio science data with astrometry, however, brings observational constraints over a much longer time span and thus significantly helps maintaining low uncertainty levels (Fig. \ref{fig:errors_europa_all_full_period}). The formal errors in the moons' tangential positions obtained from the complete dataset indeed tend to asymptotically get closer to the astrometry solution when getting further away from the missions period (2030-2035), as the influence of JUICE and Europa Clipper diminishes. Similarly, the radial position error level does not degrade as strongly as in the radio science case. For the moon's normal position errors, which are significantly more stable upon long-term propagation, adding astrometry has a smaller effect. 

To analyse the long-term error propagation further, Fig. \ref{fig:averaged_errors} displays the state uncertainty levels for all moons, averaged over the period 1890-2030 (after backward propagation). In particular, the radio science-only solution errors are extremely high for the tangential positions (as in Fig. \ref{fig:errors_europa_full_period}), and adding astrometry can reduce them by more than one order of magnitude for Io, Europa, and Ganymede. Callisto is however a notable exception: as the moon is not in resonance, the low uncertainty levels reached during the JUICE and Clipper missions remain relatively stable upon propagation. The combined solution is thus closer to the radio science case, and the astrometry dataset does not provide any significant improvement. It should however be noted that various additional perturbations and uncertainty sources, such as gravitational perturbations by the inner moons and asteroids (see detailed discussion in Section \ref{sec:conclusion}), would also affect the long-term error propagation and deteriorate the accuracy of Callisto's solution.

\subsubsection{Influence on the tidal dissipation parameters} \label{sec:resultsTidalDissipation}

\begin{table}[tbp]
	\caption{Formal uncertainties in $1/Q$ of Io and Jupiter.}
	\label{tab:invQ}
	\centering
	\begin{tabular}{l c c }
		& $1/Q_\mathrm{Jupiter}$ [-] & $1/Q_\mathrm{Io}$ [-] \\ \hline
		\textbf{Astrometry} &  & \\
		& $2.5\cdot10^{-6}$ & $1.7\cdot10^{-2}$ \\ \hline
		\textbf{Radio science} & & \\
		MONTE & $1.3\cdot10^{-6}$ & $1.0\cdot10^{-3}$ \\
		Tudat & $1.5\cdot10^{-6}$ & $3.0\cdot10^{-3}$ \\ \hline
		\textbf{Combined} & & \\
		MONTE & $3.0\cdot10^{-7}$ & $7.0\cdot10^{-4}$ \\
		Tudat & $7.3\cdot10^{-7}$ & $1.9\cdot10^{-3}$ \\ \hline
	\end{tabular}
\end{table} 

The formal errors obtained for $1/Q_{\mathrm{Jupiter}}$ and $1/Q_{\mathrm{Io}}$ in the three different configurations are reported in Table \ref{tab:invQ}. It should be noted that Tudat and MONTE provide different uncertainties for the radio science solution. While the results are surprisingly very close for $\sigma(1/Q_\mathrm{Jupiter})$ in the radio science-only case, this does not reflect the behaviour obtained for other parameters. The error in $1/Q_\mathrm{Io}$ obtained with Tudat is indeed three times larger than the one provided by MONTE, and the moons' state uncertainties (not showed in Table \ref{tab:invQ}) show similar differences (factor two to three, depending on the considered moon and direction). These discrepancies can be at least partially explained by differences in the tracking and estimation setups used for the joint JUICE-Clipper analysis, as mentioned in Section \ref{sec:parameters}. Differences between the two software can also contribute to the observed disparity (see Section \ref{sec:setups}).

Finally looking at the combined solution, the uncertainties in $1/Q_\mathrm{Jupiter}$ are a factor two to four smaller than for the radio science case, depending on the software and estimation setup. We also obtained a consistent 30-35\% improvement for $1/Q_\mathrm{Io}$ with both MONTE and Tudat. With respect to the current, astrometry-based solution, the combined solution actually represents  an order of magnitude improvement. Overall, the attainable uncertainty reduction in both $1/Q_\mathrm{Jupiter}$ and $1/Q_\mathrm{Io}$ thus seems significant. This improvement could be anticipated from the long-term propagation of the moons' state solutions shown in Fig. \ref{fig:errors_moons_full_period}. Combining radio science and astrometry indeed strongly reduces the uncertainty in the along-track direction (by more than one order of magnitude). This is crucial for the determination of the moons' tidal dissipation parameters, as the tidal effects are mostly detectable from the moons' orbits through the secular change in mean motion that they cause.  Those results, even if obtained in a simplified, (partially) simulation-based setup, indicate that adding astrometry to JUICE and Clipper data is a promising approach to better constrain tidal dissipation effects in the Jovian system.

It is worth noting that including astrometry in the solution also slightly reduces the (high) correlation between Jupiter's and Io's tidal dissipation parameters. Taking the Tudat setup as an example, the 92\% correlation between $1/Q_\mathrm{Jupiter}$ and $1/Q_\mathrm{Io}$ when relying on radio science solely is brought down to about 87\% in the combined case. This still represents a $\sim$60\% improvement in the solution departure from full correlations (8\% with radio science to 13\% with both radio science and astrometry).

\subsection{Contribution of different astrometric observable types} \label{sec:dataSetsContributions}

\begin{table*}[ht!]
	\caption{Formal uncertainties in $1/Q$ of Io and Jupiter.}
	\label{tab:dataSetsContributions}
	\centering
	\begin{tabular}{l l l| c | c c | c c}
		\multicolumn{3}{l|}{\textbf{Dataset}}  &\multicolumn{1}{l|}{\textbf{Astrometry}} &  \multicolumn{2}{c|}{$\mathbf{1\sigma}$ \textbf{[-]}} & \multicolumn{2}{c}{\textbf{Contribution astrometry }} \\ 
		\multicolumn{3}{l|}{radio science +} &\multicolumn{1}{l|}{\textbf{observations} [-]} &  $\mathbf{1/Q_\mathrm{Jupiter}}$ & $\mathbf{1/Q_\mathrm{Io}}$ & $\mathbf{1/Q_\mathrm{Jupiter}}$ & $\mathbf{1/Q_\mathrm{Io}}$ \\ \hline
		
		& \multicolumn{2}{l|}{no astrometry} & 0 & $1.5\cdot10^{-6}$ & $3.0\cdot10^{-3}$ & - & - \\
		& \multicolumn{2}{l|}{all astrometry} & 14 454 & $7.3\cdot10^{-7}$ & $1.9\cdot10^{-3}$ &  100\% & 100\% \\ \hline
		
		& \multicolumn{2}{l|}{classical astrometry} & & & & & \\
		& & all & 12 073 & $9.8\cdot10^{-7}$ & $2.3\cdot10^{-3}$ & 67\% & 64\% \\
		& & before 1960 & 2 473 & $1.2\cdot10^{-6}$ & $2.3\cdot10^{-3}$& 39\% & 64\% \\ 
		& & after 1960 & 9 600 & $1.3\cdot10^{-6}$ & $2.8\cdot10^{-3}$ & 26\% & 18\% \\
		
		& \multicolumn{2}{l|}{mutual phenomena} & 2 043 & $1.3\cdot10^{-6}$ & $2.9\cdot10^{-3}$ & 26\% & 9\% \\
		& \multicolumn{2}{l|}{ground-based radar} & 22 & $1.3\cdot10^{-6}$ & $2.8\cdot10^{-3}$ & 26\% & 18\% \\
		& \multicolumn{2}{l|}{stellar occultation} & 5 & $1.4\cdot10^{-6}$ & $2.9\cdot10^{-3}$ & 13\% & 9\% \\\hline
		
		& \multicolumn{2}{l|}{all Io observations} & 3 814 & $9.3\cdot10^{-7}$ & $2.4\cdot10^{-3}$ & 74\% & 55\% \\ \hline
		
		& \multicolumn{2}{l|}{future astrometry} & 4 877 & $1.5\cdot10^{-6}$ & $3.0\cdot10^{-3}$ & - & - \\ \hline
	\end{tabular}
	\tablefoot{The formal uncertainties are obtained from radio science simulated data combined with various subsets of the astrometric observations. The relative contributions of each data subsets are also provided. They are expressed as fractions of the total improvement achieved by adding all astrometric observations to the radio science only solution.}
\end{table*}

To further analyse which observations most effectively contribute to reducing the uncertainties in $1/Q_\mathrm{Jupiter}$ and $1/Q_\mathrm{Io}$, we ran simulations including only subsets of the available astrometry data to the simulated radioscience data. We first considered each type of observations independently. In addition to radio science, we thus separately incorporated classical astrometry, space-based astrometry, mutual phenomena, radar data, and stellar occultations. The resulting formal uncertainties are reported in Table \ref{tab:dataSetsContributions}. For each data subset, results are also provided as a percentage of the total improvement in $\sigma(1/Q_\mathrm{Jupiter})$ and $\sigma(1/Q_\mathrm{Io})$ achieved when adding all astrometric observations to radio science. We chose to show and discuss the individual contribution of different datasets based on the estimation formal errors, for which this distinction is stronger and more directly observable than for the correlations between parameters. 

When looking at the individual contribution of each data type in Table \ref{tab:dataSetsContributions}, classical astrometry has the biggest influence on the solution. More precisely, it seems that old measurements (i.e. acquired before 1960) are the most beneficial, while they only account for 20\% of all classical astrometric data, which is consistent with the discussion in Section \ref{sec:resultsStateEstimation}. As shown in Fig. \ref{fig:observations}, these observations typically show low accuracy (100s km to 1000 km). Nonetheless, they provide invaluable constraints on the long-term dynamics of the Galilean satellites and thus play a crucial role in the determination of Jupiter's and Io's tidal dissipation parameters. 

Comparatively, mutual phenomena provide a smaller improvement. They are however relatively recent observations (performed from 1973 onwards) and, as such, do not provide nearly as strong constraints as classical astrometry on the Galilean satellites' dynamics. For ground-based radar and stellar occultations, only 22 and 5 data points are respectively available. Nevertheless, their contribution to the estimation of $1/Q_\mathrm{Jupiter}$ and $1/Q_\mathrm{Io}$ is not negligible. While these are also very recent observations (starting in the 1990s and late 2010s for radar and stellar occultations, respectively), both measurement techniques demonstrate exceptional accuracy \citep[see Fig. \ref{fig:observations},][]{brozovic2020,morgado2019,morgado2022}, explains why they still provide a meaningful contribution to the solution. While ground-based radar capabilities are strongly reduced by the loss of Arecibo, these results provide strong motivation to continue observing future stellar occultation events.

The space-based astrometry data acquired during the Galileo and Voyager missions were found to not noticeably contribute to the determination of $1/Q_\mathrm{Jupiter}$ and $1/Q_\mathrm{Io}$, such that the solution including these observations is identical to the radio science only case (thus not reported in Table \ref{tab:invQ}). This result follows from the limited accuracy of the Voyager and Galileo data, but it does not reflect the quality and contribution of space-based astrometry in general. For our particular case of the Galilean satellites, the accuracy of the upcoming space astrometry data from the JUICE and Europa Clipper missions is expected to be closer to that of Cassini ISS observations, which have been proven invaluable in ephemerides and tidal dissipation studies for the Saturnian system \citep{lainey2017,lainey2019,lainey2020}.

We then considered only observations of Io, without discriminating between different types of astrometric measurements. As shown in the last row of Table \ref{tab:dataSetsContributions}, this Io-only dataset can already account for about 74\% and 55\% of the total improvement attainable when adding all astrometric observations to radio science (for  $\sigma(1/Q_\mathrm{Jupiter})$ and $\sigma(1/Q_\mathrm{Io})$, respectively). The significance of this result is strengthened by the fact that Io's data points only represent about 26\% of the entire astrometry set. This confirms that ground-based observations of Io most efficiently complement the radio science dataset, and alleviate JUICE and Clipper's lack of direct constraints on Io's dynamics. As discussed in Section \ref{sec:synergies}, Io's observations are thus a crucial aspect of the strong synergy between the radio science and astrometry.

We also quantified the contribution of potential future astrometric observations, simulated between 2023 and the beginning of JUICE's and Europa Clipper's Jovian tours (see Section \ref{sec:simulatedAstrometry}).  These additional ground-based data could however not help reducing the estimated uncertainties further, for neither Io's and Jupiter's tidal dissipation parameters nor the Galilean moons' states (see Table \ref{tab:invQ}). The added value of future astrometric observations, whose accuracy cannot compete with that of radio science measurements, directly suffers from their temporal proximity with the JUICE and Europa Clipper missions. This could be foreseen looking at formal uncertainties predicted for the radio science only solution close to the missions' period: they are indeed comparable or lower than the $\sim$150-200 km accuracy level expected for astrometric data in near-future observational campaigns. While crucial to properly constrain the ephemerides of the Galilean system before the arrival of the two spacecraft, acquiring new ground-based observations is thus not expected to noticeably improve the post-missions reconstruction of the Galilean moons' dynamics.

\section{Discussion and conclusions} \label{sec:conclusion}

We showed that adding decades of astrometry and radar observations to the radio science data expected from the upcoming JUICE and Europa Clipper missions helps estimating Io's and Jupiter's dissipation parameters. Uncertainties in Io's and Jupiter's tidal dissipation parameters are reduced by a factor two to four, depending on the software and simulation settings (see Section \ref{sec:resultsTidalDissipation}) It also stabilises the moons' dynamics solution which, if solely based on radio science tracking of the spacecraft, degrades rapidly outside the missions' time bounds (see Fig. \ref{fig:errors_moons_full_period}). Conversely, the radio science data from JUICE and Clipper will reduce the uncertainties in Io's and Jupiter's dissipation parameters by one order of magnitude with respect to the current, astrometry-based solution.

With respect to the rest of the astrometry dataset, Io's observations contribute the most to the joint radio science and astrometry solution. They indeed provide direct information about Io's orbital motion which are otherwise missing in the radio science tracking data, due to the absence of any flyby planned around that moon. Despite showing limited accuracy, old classical astrometry observations also proved very valuable thanks to the unique constraints they impose on the moons' long-term dynamics. 

On the other hand, we showed that near-future astrometric data potentially acquired before the spacecraft reach the Jovian system could not noticeably improve the joint estimation. The added-value of such observations is limited when radio science data are included in the estimation, as the latter then dominate the solution close to the JUICE and Europa Clipper missions period. Until such radio science measurements become available in the 2030s, astrometric data to be collected in the coming years are nonetheless still valuable. As has already been investigated in a separate study (V. Lainey, private communication), these observations will indeed help improving the moons' ephemerides' solution available before the missions start, which is a key aspect of the preparation effort. Additionally, more observation campaigns will be required after both missions end, to avoid the moons' state uncertainties rapidly deteriorating over time.

Our results rely on simulated measurements for JUICE's and Europa Clipper's radio science, and on the subsequent formal uncertainties obtained for the different estimated parameters. It is nonetheless worth noting that these formal errors likely indicate too optimistic uncertainty levels. In the particular case of the JUICE and Europa Clipper missions, unique challenges are expected to arise. The unprecedented accuracy of the radio science measurements, combined with JUICE's unique mission profile, indeed predicts meter-level determination of the moons' radial positions (even down to a few centimeters for Ganymede's during JUICE's orbital phase). For such estimation errors to actually be attainable, our dynamical models of both the moons and the spacecraft would need to reach comparable accuracy levels over relevant time scales, as discussed in e.g. \cite{fayolle2022}. The uncertainties in the Galilean moons' states and related dynamical parameters might thus be larger than predicted, which also implies that the improvement provided by astrometry could be stronger in practice. 

Among the different effects which will require significant model refinement for the JUICE and Clipper data to be exploited to their full potential, tidal dissipation mechanisms are critical to our understanding of the Galilean satellites' interiors and orbital evolution, which are key scientific objectives of both missions. In this analysis, we relied on the constant 1/Q assumption, but various approaches exist to incorporate tidal dissipation into the dynamical model (constant time lag, frequency-dependent $1/Q$, resonance locking \citep{fuller2016}, etc.). 
However, a fully consistent implementation for both the moons and the central planet, with a coherent modelling of the bodies' deformation response, remains ambiguous. A deeper analysis of the different tidal modelling strategies and their influence on the estimation solution will thus be crucial to provide robust results for tidal dissipation parameters \citep{magnanini2023}.

Additionally, some uncertainty sources in the satellites' dynamical model were neglected and should be analysed in future studies. This includes, among others, the gravitational perturbations by asteroids or by Jupiter's inner moons, whose orbital motions and masses are less accurately determined than those of the Galilean moons. The influence of the COM-COF offset should also be quantified. The possibility to exploit the combination of radio science and astrometry data to estimate this offset and thus mitigate its effect on the solution could also be explored. This would also further motivate the need for future astrometric observations during the JUICE and Europa Clipper missions, to complement the contribution of JUICE's altimeter GALA (mostly limited to the orbital phase around Ganymede).

Possible mismodelling of the spacecraft's dynamics would also indirectly affect the moons' ephemerides' solution. Accurately modelling all perturbations impacting the orbital motions of JUICE and Europa Clipper will thus be critical (e.g. manoeuvres, solar radiation pressure, errors in the High Accuracy Accelerometer (HAA) calibration, etc.). Moreover, in addition to the already mentioned potential inaccuracies in the current Jovian system model, time-variations in Jupiter's gravity field and rotation or inconsistencies in the satellites' rotation models could also affect the spacecraft orbit determination. Finally, for JUICE specifically, errors in the accelerometer calibration are particularly important and should be further analysed.

On the observations side, other datasets could be considered in future studies. In particular, radio science data from the Galileo and Juno missions are not included in our current work. The former is however not expected to significantly improve the solution \citep{magnanini2023}: the Galileo spacecraft could indeed only rely on an S-band, single frequency link due to the failure of the X-band high gain antenna. The resulting radio science measurements therefore showed relatively low accuracy compared to current missions \citep{jacobson2000,casajus2021}. On the other hand, the contribution of Juno data to the solution might suffer from its (temporal) proximity with the JUICE and Europa Clipper timelines, with the notable exception of the two flybys around Io planned for early 2024. These flybys are expected to bring invaluable information on Io's ill-constrained dynamics and should thus later be added to the joint solution for the Galilean moons. Finally, Gaia data, by refining the orbits of Jupiter's outer moons (Section \ref{sec:existingAstrometry}), could help quantifying the error in the Jovian ephemeris and mitigating its impact on the moons' solution. This is of particular interest for stellar occultations, as removing the contribution of Jupiter ephemeris would reduce their error budget to a few kilometres only (see Section \ref{sec:existingAstrometry}).

As already mentioned, our current results rely on simulated observations for the radio science side. In practice, many additional difficulties will arise when processing real radio science measurements and merging them with astrometry. Accurate dynamical modelling has already been identified as an important obstacle to a balanced ephemerides' solution for the Galilean satellites from JUICE and Europa Clipper data. Combining old astrometric observations with spacecraft radio science will make this requirement even more stringent by requiring our dynamical models to be consistent over both long- and short timescales. The appropriate weighting of extremely diverse data types and datasets, with different noise properties and accuracies, also represents a major challenge. Finally, quantifying and mitigating the  influence of the COM-COF offset will be crucial in future real data analyses.

Nonetheless, our analysis proves successful in generating a combined, global solution by relying on three different tools with distinct focusses and capabilities (moons' ephemerides or spacecraft dynamics, astrometry or radio science data). Our results show that exploiting the synergies between the different datasets can substantially improve the inversion solution, and the estimation of tidal dissipation parameters in particular. This encourages future efforts to work towards such a global solution, to fully exploit the JUICE and Clipper radio science measurements when they become available.

\section*{Acknowledgements}
This research was partially funded by ESA's OSIP (Open Space Innovation Platform)
program (M.F.), and supported by CNES, focused on JUICE (V.L.). A.M., M.Z., and P.T. are grateful to the Italian Space Agency (ASI) for financial support through the Agreement No. 2023-6-HH.0 in the context of ESA’s JUICE mission, and Agreement No. 2021-13-HH.0 in the context of NASA's Europa Clipper gravity and radio science experiments. A.M., M.Z., and P.T. moreover wish to acknowledge Caltech and the Jet Propulsion Laboratory for granting the University of Bologna a license to an executable version of the MONTE project edition software. 

\section*{Authors' contributions}
Most of the global inversion analysis was jointly performed by M.F. and A.M., with independent analyses using Tudat and MONTE, respectively, except for the contribution analysis of different data subsets (M.F.). The manuscript was written by M.F. This work was made possible thanks to the invaluable support from V.L. regarding the use of the NOE software and the incorporation of the astrometry data set in the inversion. It benefited from the supervision, insightful inputs and feedback of D.D., P.T. and M.Z. The authors would also like to thank the anonymous referee for their very useful comments. 

\bibliographystyle{apalike} 
\bibliography{references}

\begin{appendix}

\section{Propagated formal position uncertainties for the four Galilean satellites}\label{appendix}

\begin{figure*} [h!] 
	\centering
	\begin{minipage}[l]{1.0\columnwidth}
		\centering
		\subcaptionbox{\label{fig:errors_io_rs_full_period}}
		{\includegraphics[width=1.0\textwidth]{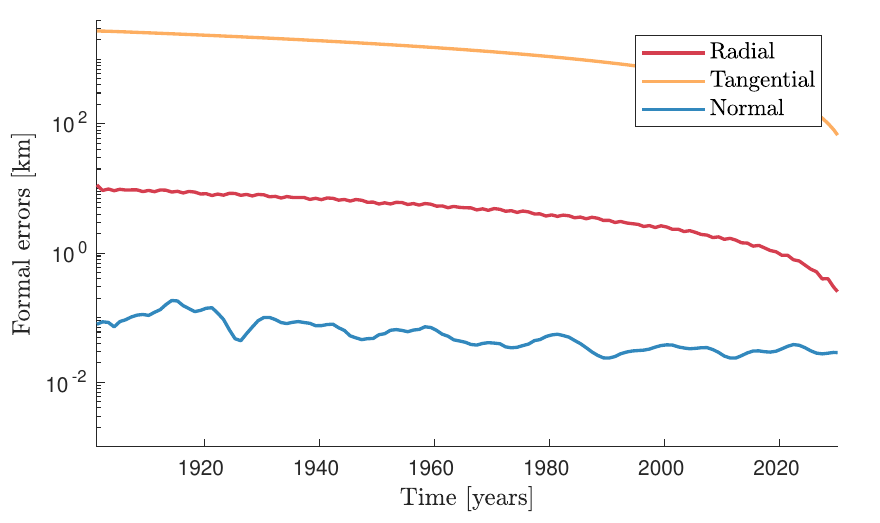}}
	\end{minipage}
	\hfill{} 
	\begin{minipage}[r]{1.0\columnwidth}
		\centering
		\subcaptionbox{\label{fig:errors_io_all_full_period}}
		{\includegraphics[width=1.0\textwidth]{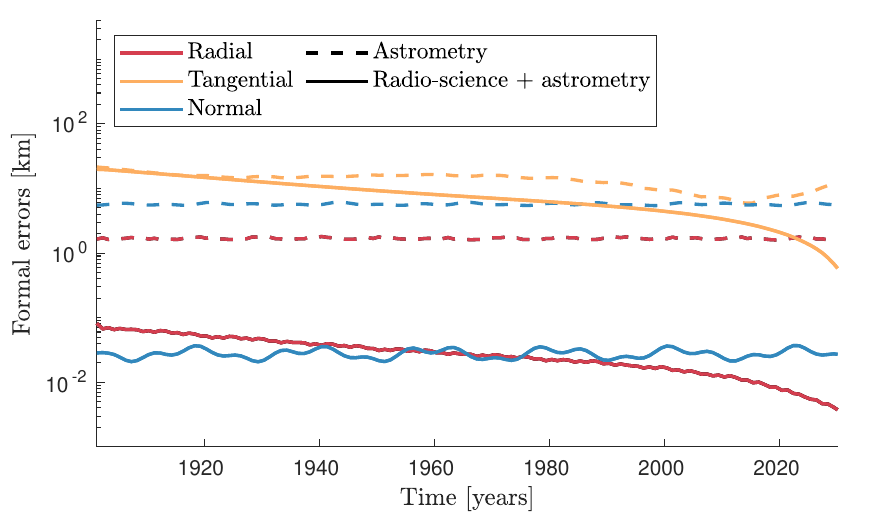}}
	\end{minipage}
	
	\begin{minipage}[l]{1.0\columnwidth}
		\centering
		\subcaptionbox{\label{fig:errors_ganymede_rs_full_period}}
		{\includegraphics[width=1.0\textwidth]{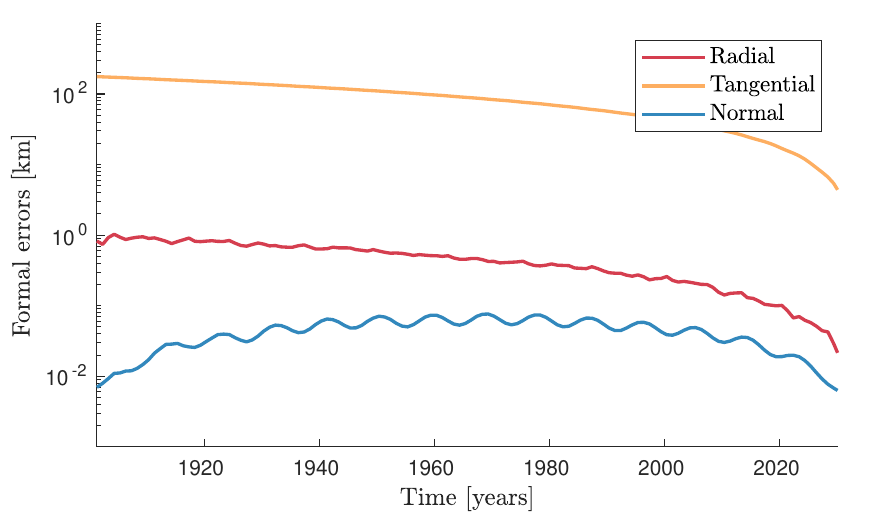}}
	\end{minipage}
	\hfill{} 
	\begin{minipage}[r]{1.0\columnwidth}
		\centering
		\subcaptionbox{\label{fig:errors_ganymede_all_full_period}}
		{\includegraphics[width=1.0\textwidth]{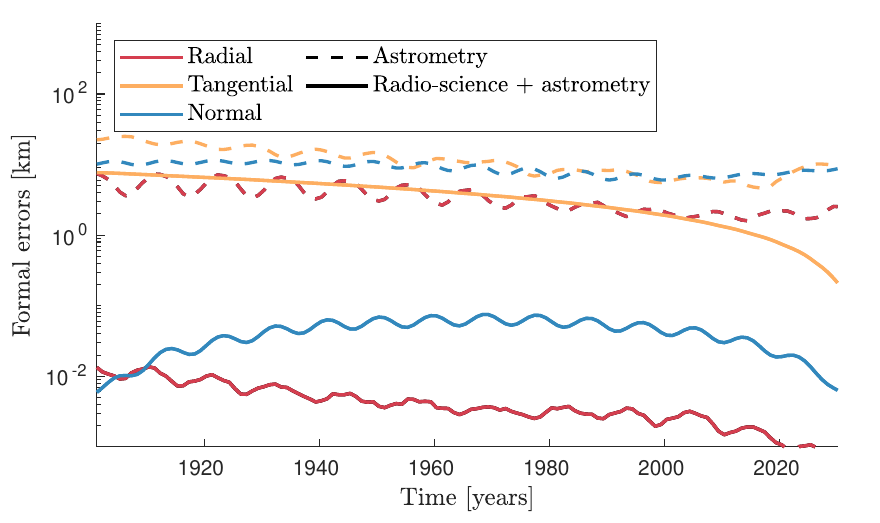}}
	\end{minipage}
	
	\begin{minipage}[l]{1.0\columnwidth}
		\centering
		\subcaptionbox{\label{fig:errors_callisto_rs_full_period}}
		{\includegraphics[width=1.0\textwidth]{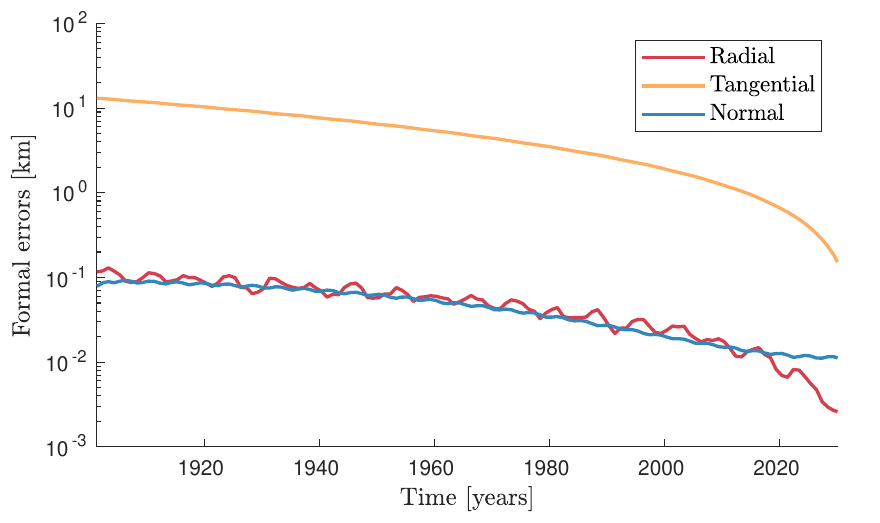}}
	\end{minipage}
	\hfill{} 
	\begin{minipage}[r]{1.0\columnwidth}
		\centering
		\subcaptionbox{\label{fig:errors_callisto_all_full_period}}
		{\includegraphics[width=1.0\textwidth]{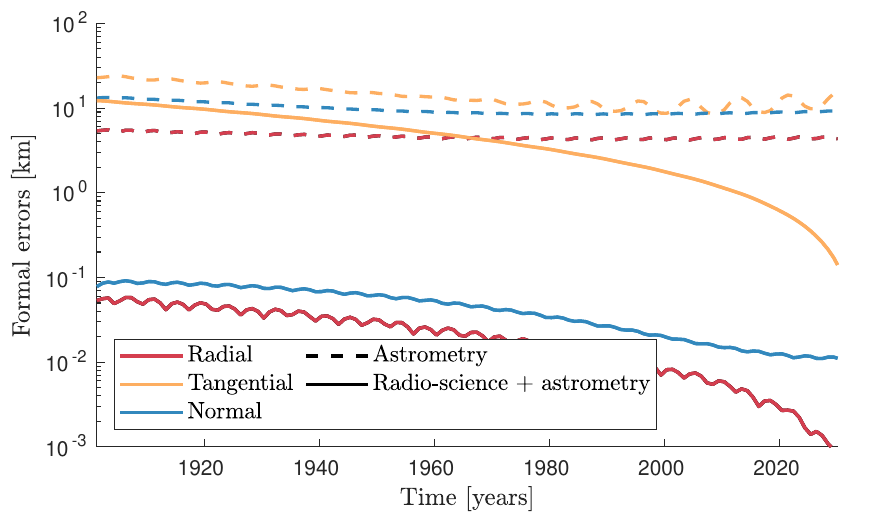}}
	\end{minipage}
	\caption{Propagated formal uncertainties in the moons' RTN positions (from 2030 to 1890) for the radio science, astrometry and combined solutions (obtained with Tudat and NOE). We used a propagation output of one point per year only and performed data smoothing over five-year windows to avoid aliasing effects (see Fig. \ref{fig:errors_europa_full_period}). The top panels present the results obtained for Io (panels a \& b), the middle panels correspond to Ganymede (panels c \& d), and the bottom ones to Callisto (panels e \& f). The left hand side panels (a, c, e) always display the radio science only solutions, while the right hand side panels (b, d, f) show both the astrometry-only and combined solutions.}
	\label{fig:errors_moons_full_period}
\end{figure*}

This appendix presents the formal position uncertainties propagated from 2033, in the middle of the JUICE and Europa Clipper science tours, to 1890 when the first astrometric observations included in our analyses were acquired. Fig. \ref{fig:errors_moons_full_period} displays the results for the four Galilean satellites, while only those obtained for Europa are shown in Section \ref{sec:resultsStateEstimation}.

\end{appendix}

\end{document}